\documentclass[twocolumn,preprintnumbers,superscriptaddress,endnote,nofootinbib,aps,prd,floatfix]{revtex4}

\usepackage{datetime}
\usepackage{color}

\usepackage{footmisc,multirow}
\usepackage{subfigure}
\usepackage{amsmath}

\usepackage{graphicx}

\renewcommand{\d}{{\text{d}}}
\newcommand{\be}{\begin{eqnarray*}}
\newcommand{\ee}{\end{eqnarray*}}
\newcommand{\gl}[1]{(\ref{#1})}

\newcommand{\bee}{\begin{eqnarray}}
\newcommand{\eee}{\end{eqnarray}}
\newcommand{\beeq}{\begin{equation}}
\newcommand{\eeeq}{\end{equation}}
\newcommand{\cp}{${\cal{CP}}$}
\newcommand{\gev}{{\text{GeV}}}
\newcommand{\tev}{{\text{TeV}}}
\newcommand{\ifb}{{\text{fb}}^{-1}}

\renewcommand{\vec}{\bf}

\preprint{IPPP/12/19} \preprint{DCPT/12/38}

\begin{document}

\title{Measuring Higgs ${\cal{CP}}$ and couplings with hadronic event shapes}

\begin{abstract}
  Experimental falsification or validation of the Standard Model of
  Particle Physics involves the measurement of the \cp~quantum number
  and couplings of the Higgs boson. Both {\sc{Atlas}} and {\sc{Cms}}
  have reported an SM Higgs-like excess around $m_H=125$~GeV. In this
  mass range the \cp~properties of the Higgs boson can be extracted
  from an analysis of the azimuthal angle distribution of the two jets
  in $pp\to Hjj$ events. This channel is also important to measure the
  couplings of the Higgs boson to electroweak gauge bosons and
  fermions, hereby establishing the exceptional role of the Higgs
  boson in the Standard Model. Instead of exploiting the jet angular
  correlation, we show that hadronic event shapes exhibit substantial
  discriminative power to separate a ${\cal{CP}}$ even from a \cp~odd
  Higgs. Some event shapes even show an increased sensitivity to the
  Higgs \cp~compared to the azimuthal angle correlation.  Constraining
  the Higgs couplings via a separation of the weak boson fusion and
  the gluon fusion Higgs production modes can be achieved applying
  similar strategies.
\end{abstract}

\author{Christoph Englert} \email{christoph.englert@durham.ac.uk}
\affiliation{Institute for Particle Physics Phenomenology, Department
  of Physics,\\Durham University, DH1 3LE, United Kingdom}

\author{Michael Spannowsky} \email{michael.spannowsky@durham.ac.uk}
\affiliation{Institute for Particle Physics Phenomenology, Department
  of Physics,\\Durham University, DH1 3LE, United Kingdom}

\author{Michihisa Takeuchi}
\email{m.takeuchi@thphys.uni-heidelberg.de} \affiliation{Institute for
  Theoretical Physics, Heidelberg University, 69120 Heidelberg,
  Germany}

\maketitle


\section{Introduction}
\label{sec:intro}
Experimental searches for the Standard Model (SM) Higgs
boson~\cite{orig} performed by {\sc{Atlas}} and {\sc{Cms}}
\cite{HiggsATLAS,HiggsCMS}, based on a combination of luminosities of
up to about 2.3~fb$^{-1}$ per experiment \cite{Rol}, exclude a SM-like
Higgs boson between 141~GeV and 476~GeV at 95\%~confidence level
(CL). By the end of December 2011 both {\sc{Atlas}} and {\sc{Cms}}
updated the Higgs search using the entirely available data set,
refining the analyses with integrated luminosities of up to
5~fb$^{-1}$~\cite{ATLASCouncil, CMSCouncil,moriond}, depending on the
channel. This allowed to raise the lower Higgs mass bound from LEP2 of
114.4~GeV~\cite{LEP2} to 117.5~by {\sc{Atlas}}. By now the Higgs is
excluded at 95\% CL from 129~(127.5)~GeV to 539~(600)~GeV by
{\sc{Atlas}} ({\sc{Cms}}). Strong bounds as low as fractions
$\sigma/\sigma^{\rm{SM}}\lesssim 0.3$ for some Higgs mass ranges have
been established. However, both {\sc{Atlas}} and {\sc{Cms}} have also
presented tantalizing hints of a $m_H\simeq 125$~GeV Higgs boson, with
local significances of 2.5$\sigma$ and 2.8$\sigma$
respectively. Together with the recently reported 2.2$\sigma$ excess
from updated searches by the Tevatron experiment~\cite{tevat}, the
hints for a light Higgs around this particular mass seem to
consolidate and various new physics interpretations of the excess have
already been considered in Refs.~\cite{bsm1,bsm2,bsm3,bsm4}. While a
$5\sigma$ discovery could be achieved in the near future, all
properties of this newly discovered state other than its mass are
going to be rather vaguely known due to limited statistics (see
{\emph{e.g.}}  Ref.~\cite{Giardino:2012ww}). The question of wether we
indeed observe the SM Higgs can only be addressed with higher
luminosity and larger center of mass energy.

A crucial step towards a further validation of the SM Higgs sector
after the discovery of the resonance is the determination of its spin,
its ${\cal{CP}}$ quantum number and its couplings to fermions and
gauge bosons. In fact, because the observed resonance seems to decay
into photons, the Landau-Yang theorem \cite{landauyang,Ellis:2012wg}
excludes the resonance to be a spin-1 particle\footnote{However,
  spin-0 or higher spin states are not excluded and an experimental
  validation is desirable.}. This leaves the measurement of the
resonance's couplings and ${\cal{CP}}$ the theoretically most
interesting ones.

In the SM, the Higgs boson is the (indispensable) remnant of the
$SU(2)$-doublet Higgs field after spontaneous symmetry breaking. To
establish that a single Higgs field is responsible for the generation
of fermion and electroweak-gauge-boson masses, eventually, the
couplings of the Higgs boson to all SM particles have to be measured
accurately. The major production processes of a light Higgs boson at
the LHC are the gluon fusion (GF) \cite{GF} and the weak boson fusion
(WBF) \cite{Rainwater:1998kj,wbf} channels. The GF channel is induced
by heavy fermion loops connecting the initial state gluons with the
Higgs boson, while the WBF channel relies on the large Higgs coupling
to electroweak gauge bosons to produce the Higgs in association with
two tagging jets. When extracting its couplings from data, the
production of the Higgs boson and its decay cannot be treated
independently \cite{sfitter}: The observed number of Higgs bosons
depends on the coupling responsible for Higgs boson production $g_p$
and the size of the coupling which dials the Higgs decay into a
specific final state $g_d$, so schematically we observe
\begin{equation}
  \label{eq:1}
  \sigma_p \cdot {\rm{BR}}_d \sim g_p^2 \frac{g_d^2}{\Gamma_H}\,.
\end{equation}
Note that even if $g_p = g_d$ the total width of the Higgs boson
$\Gamma_H$ is sensitive to all Higgs couplings, but a direct
measurement of $\Gamma_H$ is not possible at hadron colliders due to
systematics.

A channel, which is phenomenologically well-suited to study
longitudinal gauge boson scattering \cite{longitud} and Higgs
couplings to electroweak gauge bosons is $pp\to\text{Higgs+2 jets}$
(with subsequent Higgs decay). In this channel, it is particularly
difficult to separate gluon fusion from the weak
boson fusion contribution since both production modes exhibit similar
cross sections for typical event selection cuts
\cite{hwg,kinem}. Because $g_p \simeq g_{p,\text{GF}} +
g_{p,\text{WBF}}$ in Eq.~\gl{eq:1}, the uncertainties of different
Higgs couplings obtained from experimental analyses in this channel
are correlated and the extraction of the individual couplings becomes
challenging \cite{sfitter}.

For a 125~GeV SM Higgs-like resonance, we have to face the
phenomenological impediment that standard ${\cal{CP}}$ analyses
\cite{cp} of the so-called gold-plated final state $H\to ZZ \to 4\ell$
\cite{Bredenstein:2006rh}, which employ strategies closely related to
the one proposed by Cabibbo and Maksymowicz in the context of kaon
physics \cite{Cabibbo:1965zz,Dell'Aquila:1985ve,Collins:1977iv} are
statistically limited even at $\sqrt{s}=14$ TeV. Instead, the
jet-azimuthal angle correlation in Higgs+2 jets events with $H\to
\tau^+\tau^-$ has been put forward as an excellent probe of the
${\cal{CP}}$ nature of the Higgs boson in series of seminal papers
\cite{Plehn:1999xi,Plehn:2001nj,Rainwater:1998kj}. Since then a lot of
effort has been devoted to theoretical and phenomenological
refinements of this important channel. These range from precise (fixed
higher order QCD) predictions of the contributing signal and
background processes \cite{Campbell:2006xx,backgroundtt,Moch:2008ai,
  backgroundz,kinem} over resummation
\cite{Andersen:2008u,Andersen:2008} to the generalization to the other
important final state for a light Higgs, $H\to WW$
\cite{Klamke:2007cu,ken}. Only recently, the $pp\to Hjj \to
\tau^+\tau^-jj$ channel was studied for the first time at the LHC to
derive bounds on the SM Higgs boson production cross section
\cite{Chatrchyan:2012vp}. This impressively demonstrates that
experimental systematics in this important channel are well under
control, already now with early data.

The azimuthal angle correlation of the tagging jets as a
${\cal{CP}}$-discriminative observable can be rephrased in the
following way: Once the Higgs is identified, the hadronic energy flow
of the event depends on the ${\cal{CP}}$ quantum number of the
produced Higgs. The correlation of in the azimuthal angle should also
be reflected in the global structure of softer tracks, which do not
give rise to resolved jets. It is precisely the hadronic energy flow
which is captured by event shape observables in theoretically
favorable way \cite{evtshapes}, turning them into natural candidates
to be considered among the ${\cal{CP}}$-discriminative observables in
the context of \cp~analyses. From a perturbative QCD point of view,
the phenomenology of event shapes~\cite{evtshapes} possesses a number
of advantages over ``traditional'' jet-based observables. In
particular, provided that the observables are ``continuously global''
\cite{Dasgupta:2002dc,resumevtshapes}, they can be resummed to NLL
beyond the leading color approximation. Therefore, event shapes offer
a good theoretical handle to potentially reduce perturbative
uncertainties.

We organize this work in the following way: Sec.~\ref{sec:shapes}
briefly reviews the hadronic event shape and the $\Delta\Phi_{jj}$
observables, which we consider in the course of this paper. We outline
the details of our analysis in Sec.~\ref{sec:analysis}. We discuss the
sensitivity of event shapes in ${\cal{CP}}$ analyses of Higgs+2 jets
events in Sec.~\ref{sec:results}, where we also investigate the
possibility to distinguish WBF from GF invoking the same
observables. Before we give our conclusions and an outlook in
Sec.~\ref{sec:conc}, we briefly comment on pile-up issues that can
arise in the suggested analysis in Sec.~\ref{sec:pile}.

\section{Event Shape observables and $\Delta\Phi_{jj}$}
\label{sec:shapes}
Event shapes quantify geometrical properties of the final state's
energy flow\footnote{The phenomenology and resummation of a large
  class of event shape observables at hadron colliders has recently
  been discussed in Refs.~\cite{evtshapes,resumevtshapes}.}.  An event
shape, which is well-known from QCD measurements performed during the
LEP era \cite{thrustexi,thrust} is {\bf{thrust}} $T$
\cite{Brandt:1964sa}. In its formulation in the
beam-transverse plane this observable is also meaningful at hadron
colliders,
\begin{equation}
  \label{eq:thrust}
  T_{\perp,g} = \max_{{\vec{n}}_T} \frac{ \sum_i |{\vec{p}}_{\perp,i} \cdot
    {\vec{n}}_T|}{\sum_i | {\vec{p}}_{\perp,i}|}\,.
\end{equation}
The subscript $g$ indicates that this is a continuously global
observable~\cite{resumevtshapes}. The three vectors
${\vec{p}}_{\perp,i}$ are the beam-transverse momentum components of
the particle $i$ ({\emph{i.e.}}  a {\sc{Atlas}} topocluster or a
{\sc{Cms}} particle flow object), while the sum runs over all detected
particles (typically in $|\eta_i|\leq 4.5$). In a nutshell,
$T_{\perp,g}$ measures how circularly symmetric ($T_{\perp,g} \to
2/\pi$) or how pencil-like \hbox{($T_{\perp,g}\to 1$)} an event
appears to be in the transverse plane. The vector ${\vec{n}}_T$ in the
transverse plane that maximizes Eq.~\gl{eq:thrust} is called the
transverse thrust axis.

Another event shape, familiar from $e^+e^-$ physics, which can be
straightforwardly adapted to hadron collider physics analogous to
Eq.~\gl{eq:thrust} is {\bf{thrust~minor}}
\begin{equation}
  \label{eq:thrustmin}
  T_{m,g} = \frac{ \sum_i |{\vec{p}}_{\perp,i} \times
    {\vec{n}}_T|}{\sum_i | {\vec{p}}_{\perp,i}|} \,.
\end{equation}
$T_{m,g}$ provides a measure of the energy flow in the transverse
plane perpendicular to ${\vec{n}}_T$. Both transverse thrust and
thrust minor have already been studied with early LHC
data~\cite{thrustcms}.

As already mentioned, the tagging jet azimuthal angle correlation is a
${\cal{CP}}$-discriminative observable in Higgs+2 jets
production. $\Delta\Phi_{jj}$ can be defined as the angle between all
jets $j$ with rapidity smaller and all jets with rapidity larger than
the reconstructed Higgs \cite{kinem,Andersen:2008}
\begin{equation}
  \label{eq:phijj}
  \begin{gathered}
    p^\mu_<  = \sum_{{{j\in \{\rm{jets}}} :\; y_j<y_h\}} p^\mu_j \,,  \quad
    p^\mu_>  = \sum_{{{j\in \{\rm{jets}}} :\; y_j>y_h\}} p^\mu_j \\[0.1cm]
      \Delta\Phi_{jj} = \phi(p_>) - \phi(p_<)\,.
  \end{gathered}
\end{equation}
The special role played by the tagging jets in $\Delta\Phi_{jj}$ is
best reflected in the {\bf{cone thrust minor}} event shape. Its
definition is similar to Eq.~(\ref{eq:thrustmin}), but only particles
which fall into the vicinity of two reconstructed kT jets
\cite{Catani:1993hr} with some resolution $D$ (we will assume $D=0.4$
in the following) are considered in the sum.

Typical selection cuts which are used to suppress the contributing
backgrounds often involve the requirement that the tagging jets fall
into opposite hemispheres $y_{j_1}\cdot y_{j_2}<0$ while the Higgs is
produced in the central part of the detector. Observing ${\cal{CP}}$
sensitivity in the $\Delta\Phi_{jj}$ distribution suggests that
broadening observables \cite{broadening} also carry information about
the Higgs ${\cal{CP}}$. We divide the event up according to the
transverse thrust axis
\begin{subequations}
  \label{eq:wrapbroad}
  \begin{equation}
    \begin{split}
      \text{region $D$:}\quad & {\vec{p}}_{\perp,i}\cdot {\vec{n}}_{T} > 0
      \\
      \text{region $U$:}\quad & {\vec{p}}_{\perp,i}\cdot {\vec{n}}_{T} < 0
    \end{split}
  \end{equation}
  and compute the weighted pseudorapidity and azimuthal angle
  \begin{multline}
    \eta_{X}=\frac{ \sum_{i} |{\vec{q}}_{\perp,i}|
      \, \eta_i }{
      \sum_{i} |{\vec{q}}_{\perp,i}|} \,,
    \qquad
    \phi_{X}=\frac{ \sum_{i} |{\vec{q}}_{\perp,i}|
      \, \phi_i }{
      \sum_{i} |{\vec{q}}_{\perp,i}|}\,,\\X=U,D.
  \end{multline} 
  $\eta_i$ and $\phi_i$ are the pseudorapidity and azimuthal angle of
  the vector $i$ respectively.  From these we can compute the
  broadenings of the $U$ and $D$ regions
  \begin{multline}
    B_{X}={1\over Q_{T} }\sum_{i \in X}
    |{\vec{q}}_{\perp,i}|\sqrt{(\eta_i -
      \eta_{X})^2+(\phi_i-\phi_{X})^2}\,,\\X=U,D
  \end{multline}
  where 
  \begin{equation}
    Q_{T} = {\sum_i | {\vec{q}}_{\perp,i}|}\,.
  \end{equation}
  The {\bf{central total broadening}} and {\bf{wide broadening}} are
  defined as \cite{evtshapes,thrustexi}
  \begin{equation}
    \label{eq:broadening}
    \begin{split}
      \text{central total broadening: }&
      B_{T}=B_{U}+B_{D}\,, \\
      \text{wide  broadening: }&
      B_{W}=\max\left\{B_{U},B_{D}\right\}\,.
    \end{split}
  \end{equation}
\end{subequations}

The observables Eqs.~\gl{eq:thrust},~\gl{eq:thrustmin}, and
\gl{eq:wrapbroad} do not exhaust the list of existing event shapes by
far but they are sufficient for the purpose of this work.

\section{Elements of the Analysis}
\label{sec:analysis}
\subsection{Event generation}
\subsubsection*{Signal}
Event shapes are known to be well-reproduced by matched shower Monte
Carlo programs \cite{evtshapes}. Therefore, we generate MLM-matched
\cite{mlm} scalar $Hjj$ and pseudoscalar $Ajj$ samples with
{\sc{MadEvent}} v4 \cite{madevent} in the effective $ggH$ and $ggA$
coupling approximation \cite{effhiggs}
\begin{equation} 
  \label{eq:efflag}
  {\cal{L}}={\alpha_s\over 12\pi v} HG^{a}_{\mu\nu}
  G^{a\, \mu\nu} + {\alpha_s\over 16\pi v} A G^{a}_{\mu\nu} \tilde G^{a\,
    \mu\nu}\,,
\end{equation}
where $G^{a}_{\mu\nu}, \tilde G^{a}_{\mu\nu}$ are the gluon field
strength and the dual field strength tensor, respectively, and $v$
denotes the Higgs vacuum expectation value. We subsequently shower
the events with {\sc{Pythia}} \cite{Sjostrand:2006za}.  We normalize
the event samples to the NLO QCD cross section, which we obtain by
running {\sc{Mcfm}} \cite{mcfm} for the gluon fusion contributions,
and {\sc{Vbfnlo}} \cite{vbfnlo} for the weak boson fusion
contributions. The interference effects are known to be negligible for
weak boson fusion cuts \cite{hjjint}.  Note that there is no WBF
contribution for the ${\cal{CP}}$ odd scalar $A$. Nonetheless it is
customary to analyze $Ajj$ and $Hjj$ samples for identically chosen
normalizations to study the prospects of discriminating
``Higgs-lookalike'' scenarios \cite{lykken, cphadzz}.

We find a total Higgs-inclusive normalization (considering
$\sqrt{s}=14$ TeV) of $\sigma_H=3.2~\text{pb}$. For the ${\cal{CP}}$
odd scalar we use $\sigma_A=2.1~\text{pb}$ which adopts the NLO QCD
gluon fusion $K$ factor of ${\cal{CP}}$ even Higgs production. In
Sec.~\ref{sec:HLL} we also discuss our results for identical
normalizations, which focuses on the discriminating power of different
shapes instead of a combination of shapes and different total cross
sections. The ditau branching ratio to light opposite lepton flavors
is approximately 6.2\%.

\begin{table*}[!p]
  \begin{tabular}{ c | c | c ||  c | c}
    & $t\bar t $+ jets  & $Z$+2~jets & $H$+2~jets & $A$+2~jets \\
    & $\sigma$~[fb] & $\sigma$~[fb] & $\sigma$~[fb] & $\sigma$~[fb]  \\
    \hline
    $p_{T,j}\geq 40~\gev,~|y_j|\leq 4.5,~n_j\geq 2$
    &\multirow{2}{*}{2132.46}  
   &\multirow{2}{*}{8.52}  & \multirow{2}{*}{6.21}  & \multirow{2}{*}{4.12}\\
    $p_{T,\tau}\geq 20~\gev,~|\eta_\tau|\leq 2.5\,~n_\tau=2$  & & & &\\ 
    \hline
    $m_{jj}\geq 600~\gev$ & 145.68 & 3.98 &  4.12 & 1.87\\
    \hline
    $|m_{\tau\tau}-m_H|<20~\gev,~|y_H|\leq 2.5$ & 99.86 & 2.29 & 3.99
    & 1.82 \\
    \hline
    $\exists\, j_a,j_b: y_{j_a}<y_h<y_{j_b} $ &  88.33 & 1.65 & 3.81 & 1.59
    \\
    \hline
    $b$-veto  & 5.10 & 1.65 & 3.81 & 1.59 \\
    \hline
  \end{tabular}
  \caption{\label{tab:cut-flow} Cut-flow of the analysis as described
    in Sec.~\ref{sec:selection}. For $Z+$2 jets, $H+$2 jets and $A+$2 
    jets we normalize to their NLO QCD cross section. The $t\bar{t}$ 
    production cross section we normalize to the NNLO QCD cross 
    section given in \cite{Moch:2008ai}. We neglect tau reconstruction 
    efficiencies throughout. $n_\tau$ and $n_j$ denote the
    tau and jet multiplicities, respectively.}
  \vfill
\end{table*}
\begin{figure*}[!p]
  \begin{center}
    \includegraphics[width=0.43\textwidth]{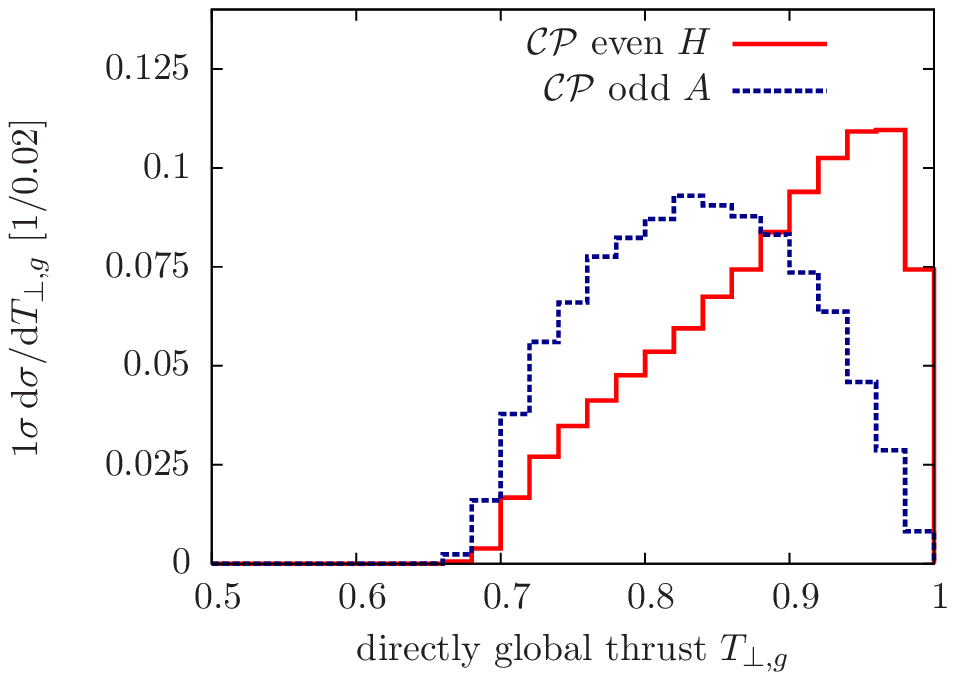}
    \hskip0.6cm
    \includegraphics[width=0.43\textwidth]{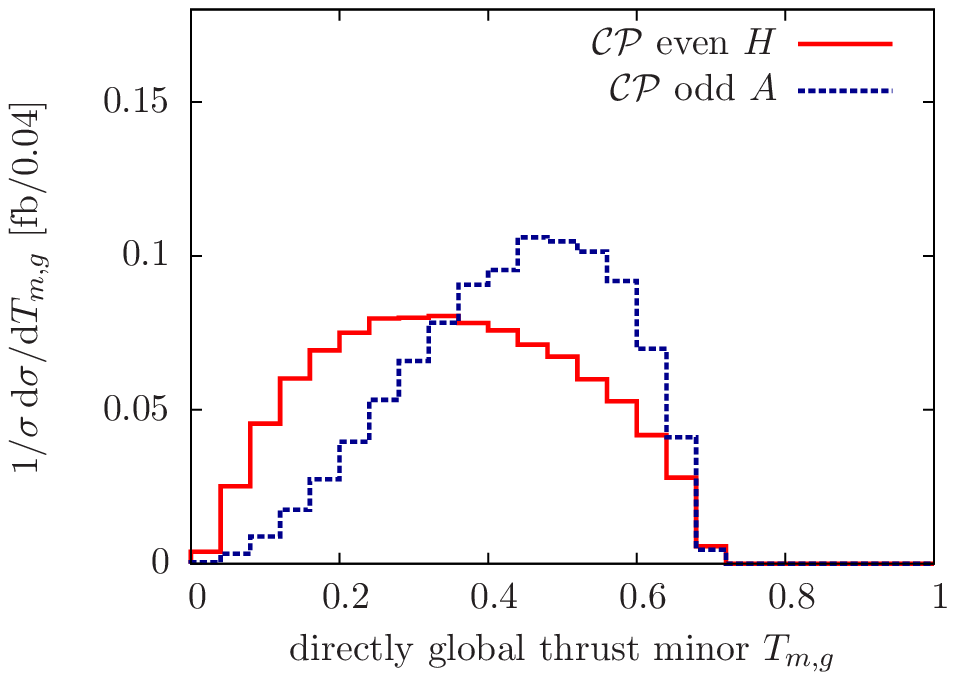}\\[0.3cm]
    \includegraphics[width=0.43\textwidth]{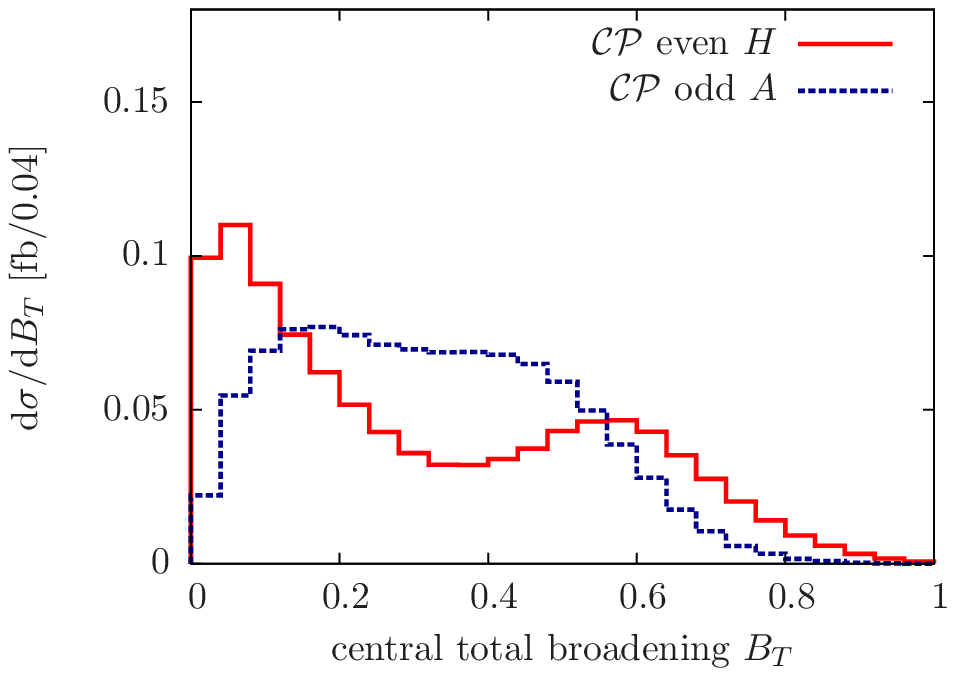}
    \hskip0.6cm
    \includegraphics[width=0.43\textwidth]{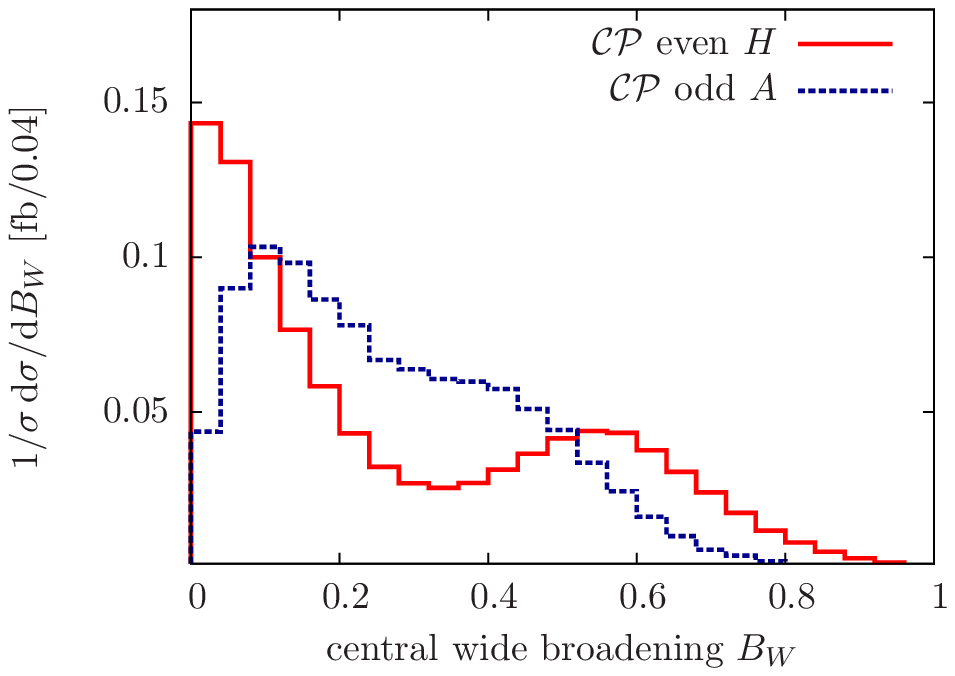}\\[0.3cm]
    \parbox{0.43\textwidth}{
      \includegraphics[width=0.43\textwidth]{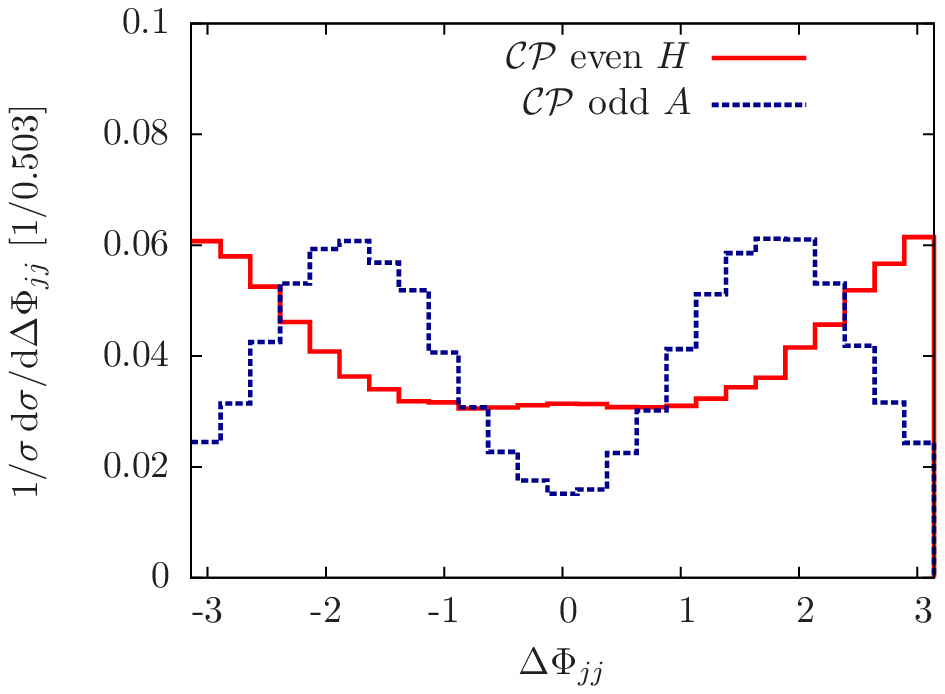}
      }
    \hskip0.6cm
    \parbox{0.43\textwidth}{
      \vspace{-1cm}
      \hspace{0.5cm}
      \parbox{0.35\textwidth}{ \caption{\label{fig:distrib} Normalized
          distributions of $\Delta\Phi_{jj}$ and of the event shape
          observables of Sec.~\ref{sec:shapes}. The cuts of
          Sec.~\ref{sec:selection} have been applied.}  
      } }
  \end{center}
\end{figure*}

\subsubsection*{Backgrounds}
We focus on the two main backgrounds to our
analysis~\cite{Plehn:1999xi}, {\emph{i.e.}} $t\bar t$+jets and $Zjj$
production, where the $Z$ boson decays to taus. We generate our
CKKW-matched \cite{ckkw} event samples with {\sc{Sherpa}}
\cite{sherpa}. We again obtain NLO QCD normalizations of the $Zjj$
sample from a combination of {\sc{Mcfm}} and {\sc{Vbfnlo}} for the QCD
and EW production modes, respectively, and find
\hbox{$\sigma(Z\to\tau^+\tau^-)=0.23$ pb}. For the $t\bar t$ sample we
extract the NNLO-inclusive $t\bar t$ $K$~factor from the cross section
$\sigma_{t \bar t}^{\text{NNLO}} = 918$~pb \cite{Moch:2008ai} in
comparison with the cross section by {\sc{Sherpa}} after
generator-level cuts $\sigma_{t \bar t}=888.27$~fb, which already
requires the tau leptons to reconstruct $m_H=125~\gev$ within
$50~\gev$.

\begin{figure}[!h]
  \begin{center}
    \subfigure[][~${\cal{CP}}$ even Higgs]{
      \includegraphics[angle=-90,width=0.4\textwidth]{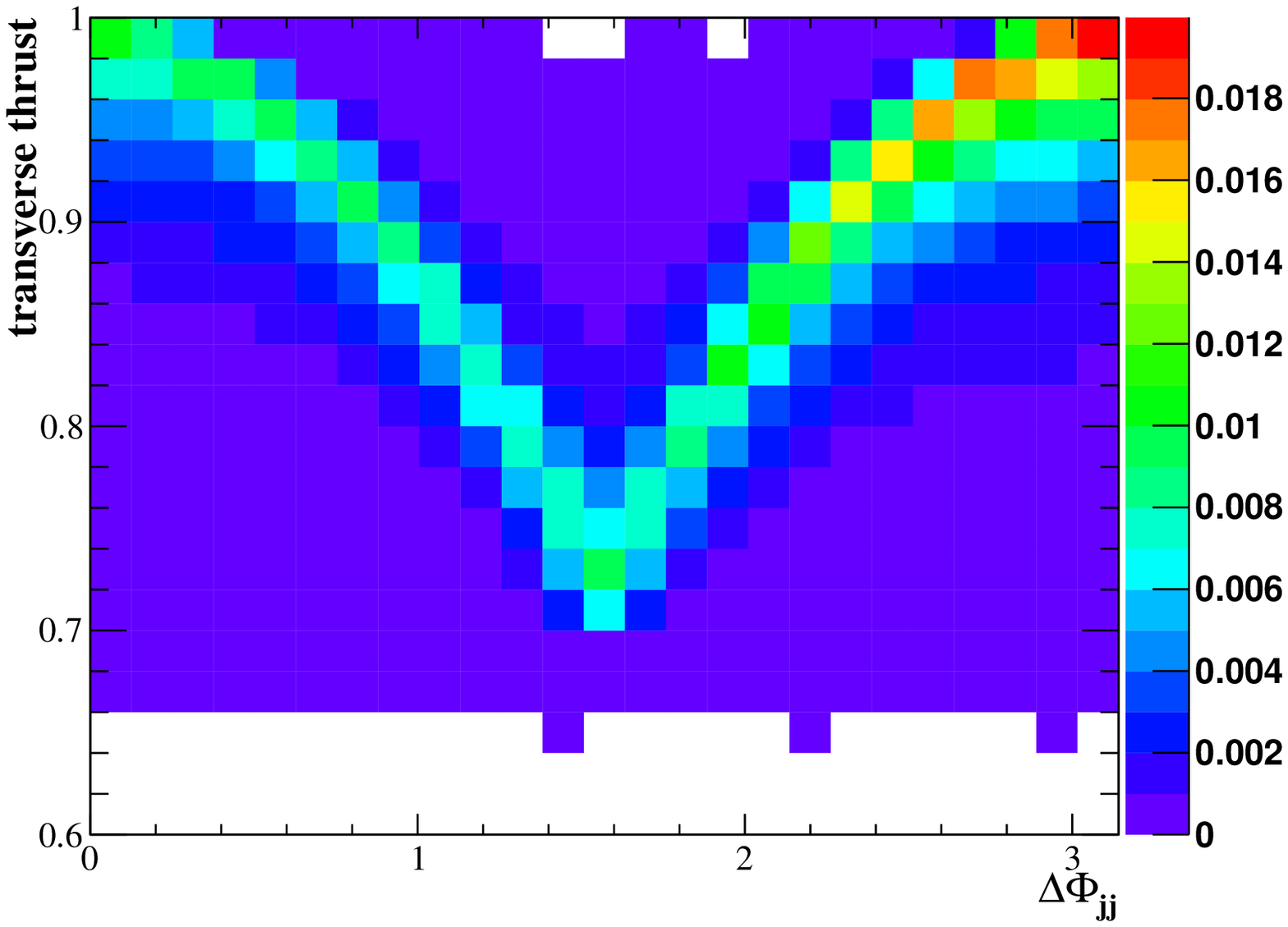}
    } \\[0.3cm]
    \subfigure[][~${\cal{CP}}$ odd Higgs]{
      \includegraphics[angle=-90,width=0.4\textwidth]{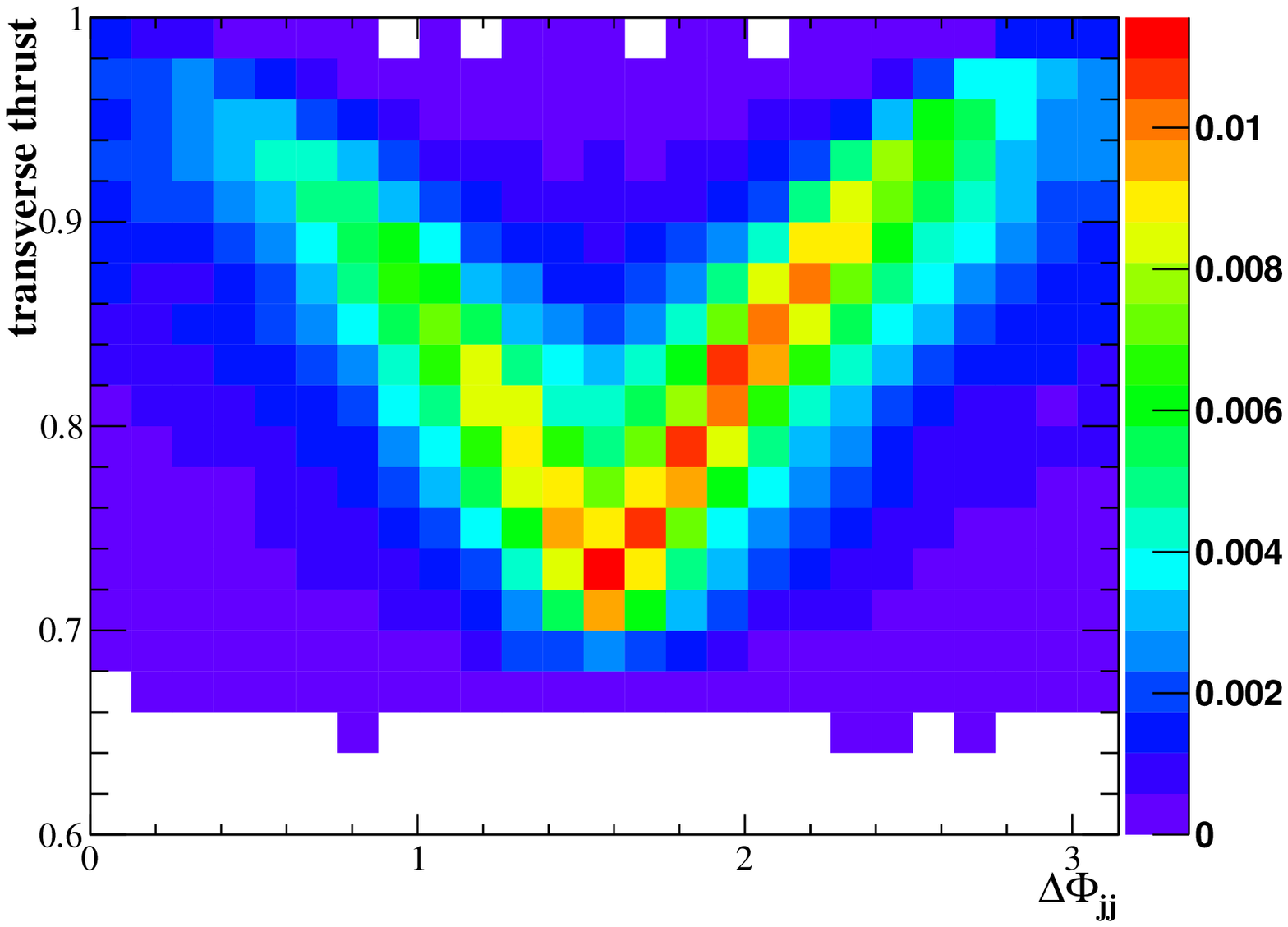}
    } 
  \end{center}
  \caption{\label{fig:corr} Correlation of the thrust event shape with
    $\Delta\Phi_{jj}$ angle as defined in Eq.~\gl{eq:phijj} in terms
    of the 2d differential probability distribution $1/\sigma\, {\d^2
      \sigma/(\d \Delta\Phi_{jj}\, \d T_{\perp,g} )}$}
  \vspace{-0.2cm}
\end{figure}
\begin{figure}[!b]
  \begin{center}
    \subfigure[][~${\cal{CP}}$ even Higgs]{
      \includegraphics[angle=-90,width=0.4\textwidth]{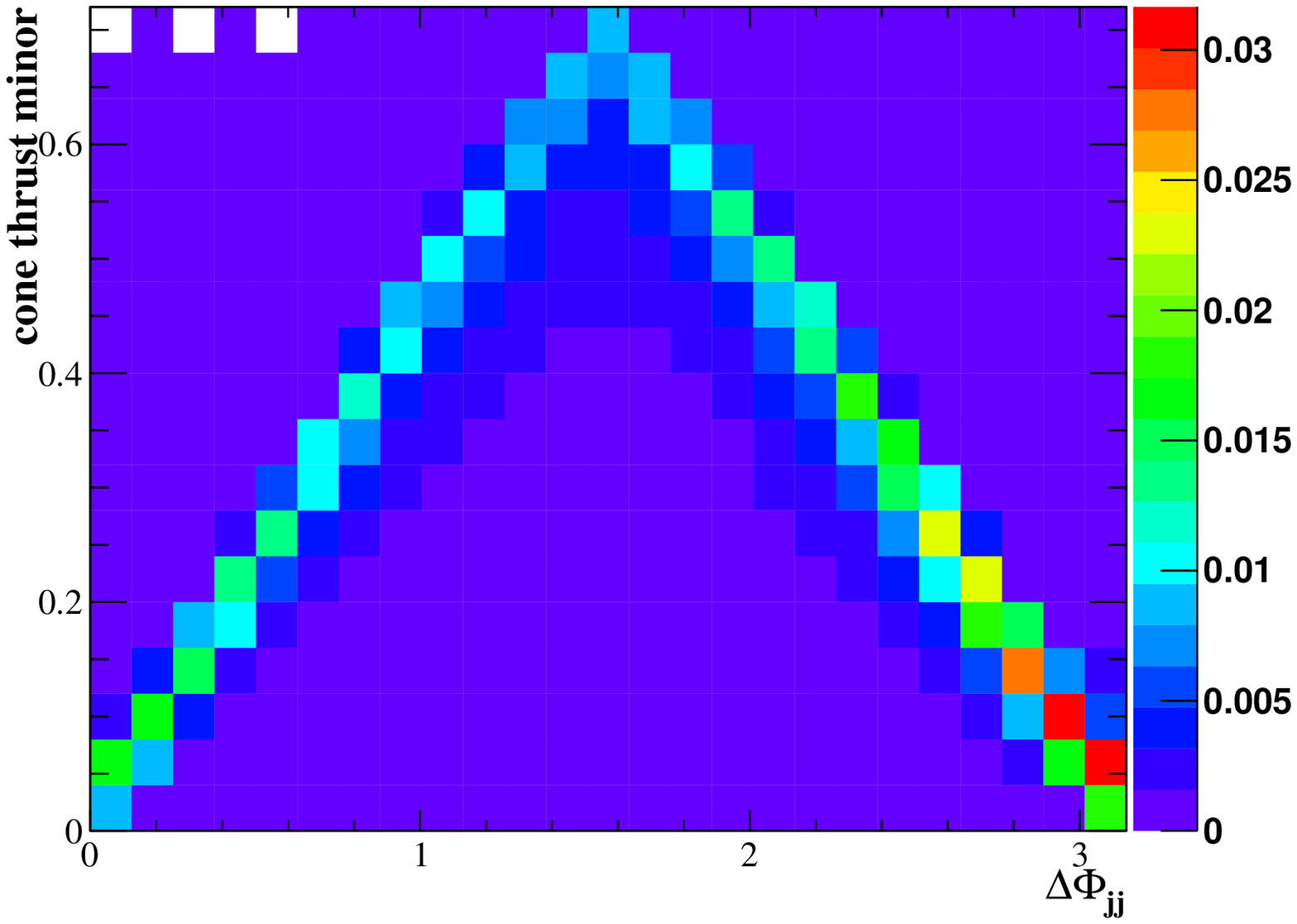}
    } \\[0.3cm]
    \subfigure[][~${\cal{CP}}$ odd Higgs]{
      \includegraphics[angle=-90,width=0.4\textwidth]{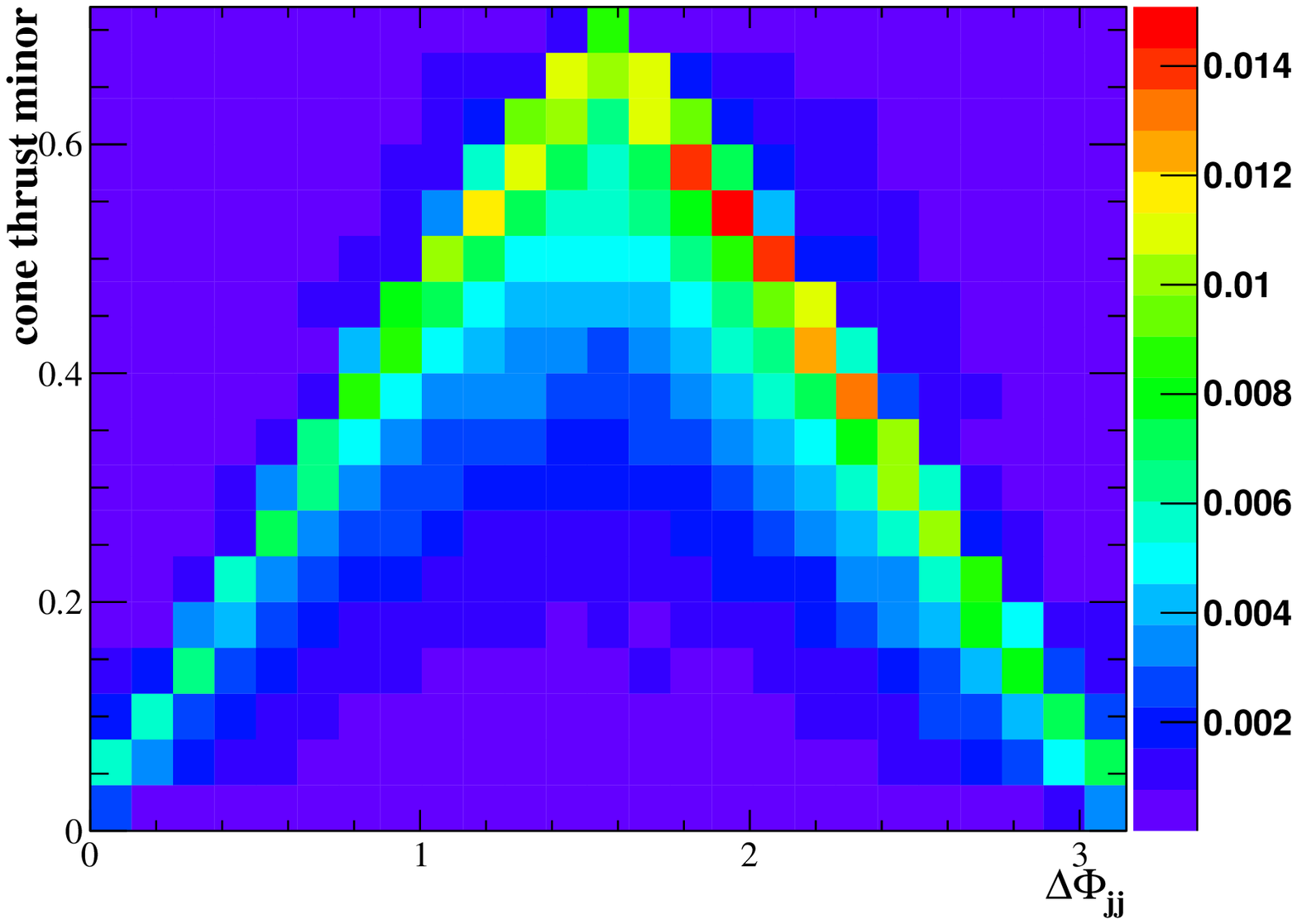}
    } 
  \end{center}
  \caption{\label{fig:corrcone} Correlation of the cone thrust minor
    event shape with $\Delta\Phi_{jj}$ angle as defined in
    Eq.~\gl{eq:phijj} in terms of the 2d differential probability
    distribution $1/\sigma\, {\d^2 \sigma/(\d \Delta\Phi_{jj}\, \d
      T_{C,m} )}$}
  \vspace{-0.2cm}
\end{figure}

\subsection{Selection Cuts and Analysis Strategy}
\label{sec:selection}
The purpose of this paper is a comparison of the ${\cal{CP}}$ and
GF/WBF discriminative power of the observables of
Sec.~\ref{sec:shapes}. The possibility to perform Higgs searches in
this channel has already been demonstrated in the literature
\cite{Plehn:1999xi,Chatrchyan:2012vp} and so we have a situation in
mind, when the Higgs is well established in this particular channel,
i.e. has a large enough significance $S/\sqrt{B}$ with reasonable
signal-to-background ratio $S/B$. Hence we apply selection strategies
and efficiencies which closely follow the parton-level analysis of
Ref.~\cite{Plehn:1999xi} to obtain an estimate of $S/B$, but our
selection should be understood as placeholder for a dedicated cut
setup. The experiments $S/B$ will hence be different, yet the impact
of $S/B$ on the observables of Sec.~\ref{sec:shapes} is identical and
a comparison is still meaningful.

We reconstruct jets with the anti-kT jet
algorithm \cite{antikt} with parameter $D=0.4$ as implemented in
{\sc{FastJet}} \cite{fastjet}. We additionally impose typical weak
boson fusion cuts to suppress the background to a manageable
level. More specifically, we require at least two jets with
\begin{subequations}
  \label{eq:selectiona}
  \begin{equation}
    p_{T,j}\geq 40~\gev,~{\text{and}}~|y_j|\leq 4.5\,,
  \end{equation}
  and the two hardest (``tagging'') jets in the event are required to have a large
  invariant mass
  \begin{equation}
    \label{eq:selectiona1}
    m_{jj}=\sqrt{(p_{j,1}+p_{j,2})^2}\geq 600~\gev\,.
  \end{equation}
\end{subequations}
After these cuts the signal is still dominated by the $t\bar t $+jets
background. This background, however, can be efficiently suppressed
with a $b$ veto from the top decay.

\begin{figure*}[!p]
  \begin{center}
    \includegraphics[width=0.43\textwidth]{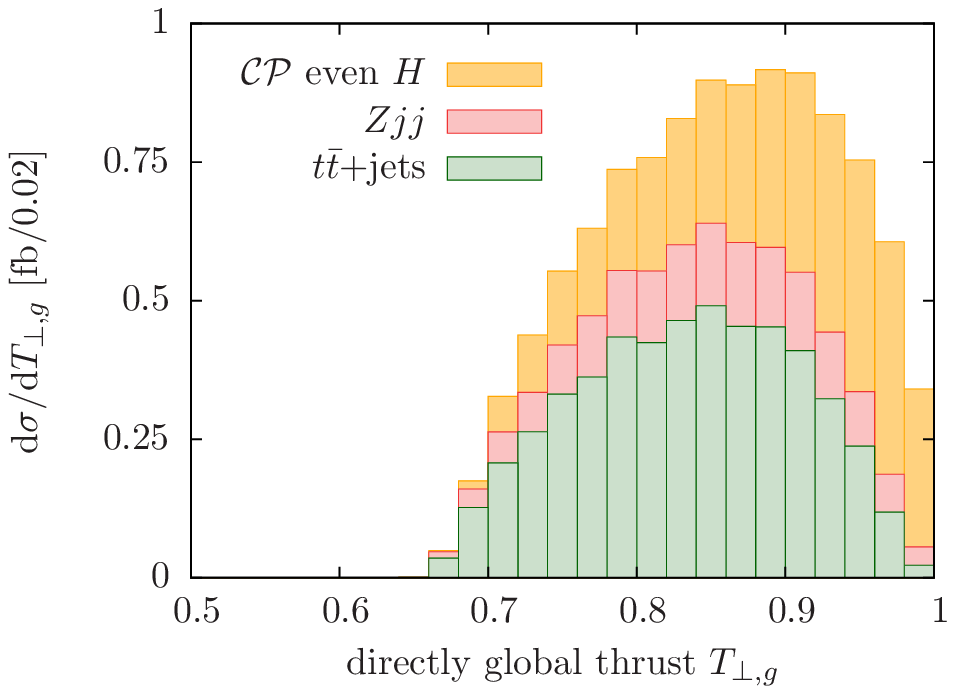}
    \hskip0.6cm
    \includegraphics[width=0.43\textwidth]{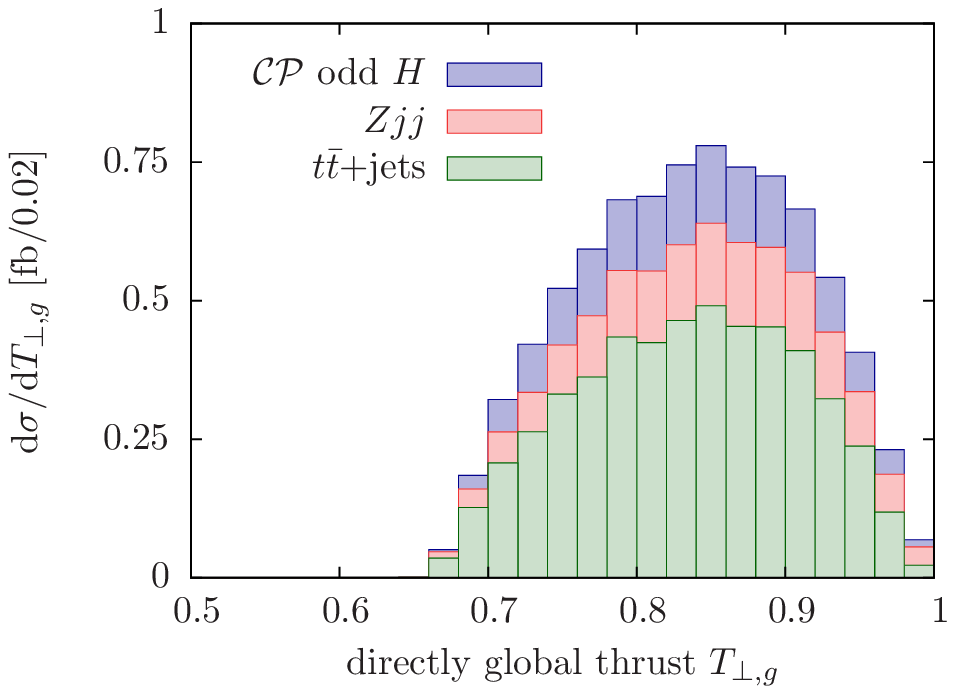}\\[0.5cm]
    \includegraphics[width=0.43\textwidth]{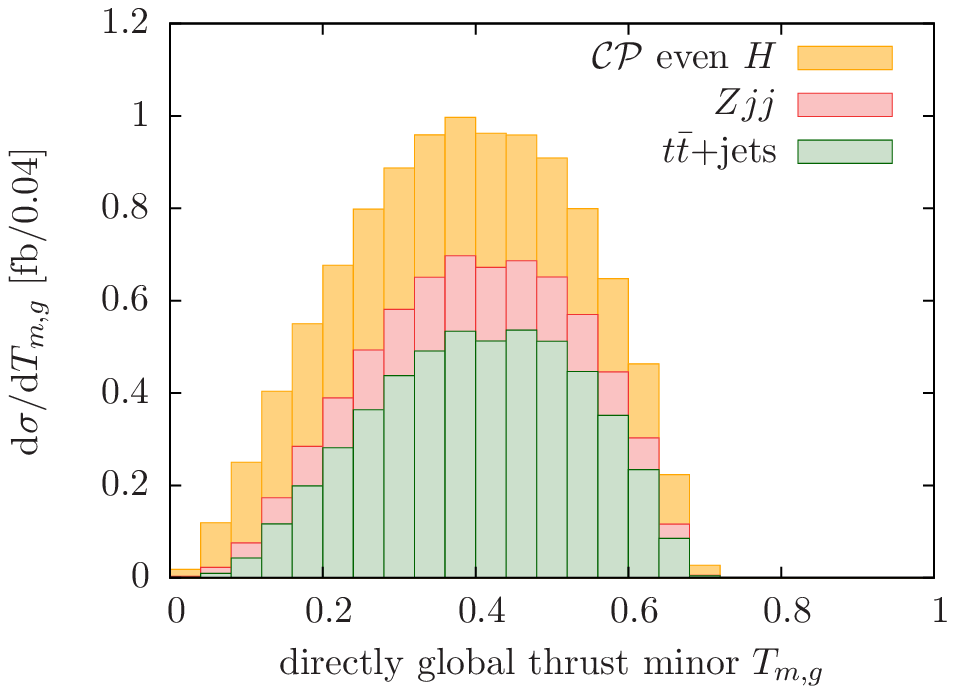}
    \hskip0.6cm
    \includegraphics[width=0.43\textwidth]{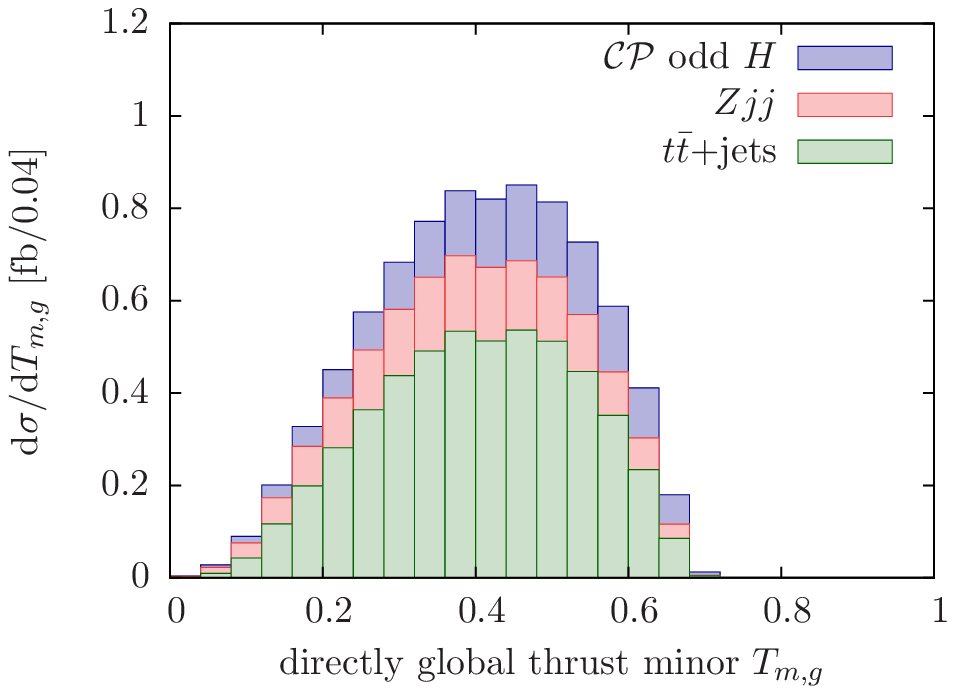}\\[0.5cm]
    \includegraphics[width=0.43\textwidth]{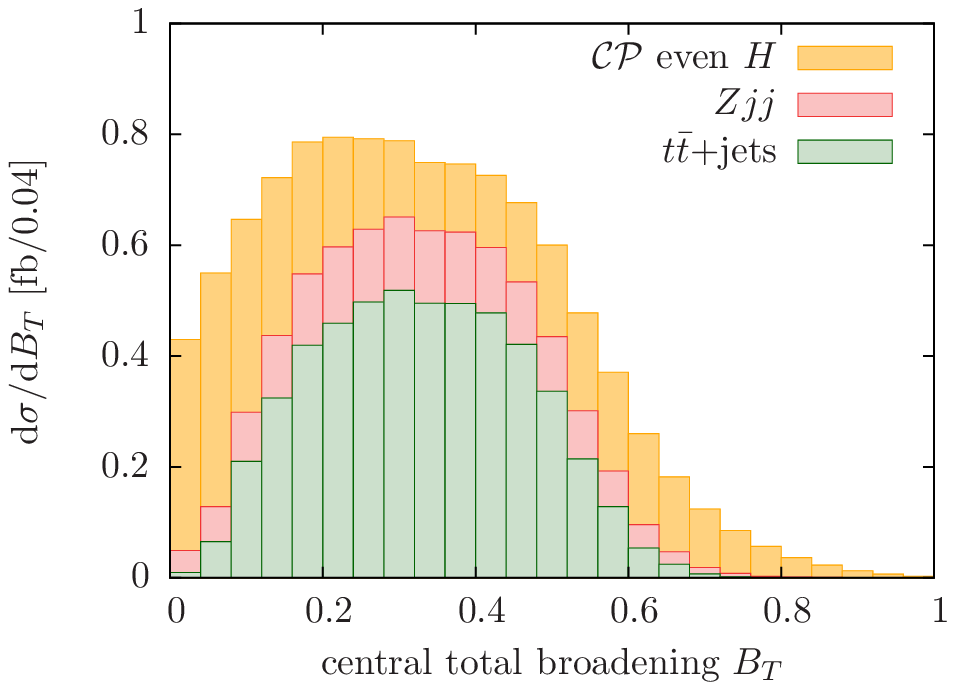}
    \hskip0.6cm
    \includegraphics[width=0.43\textwidth]{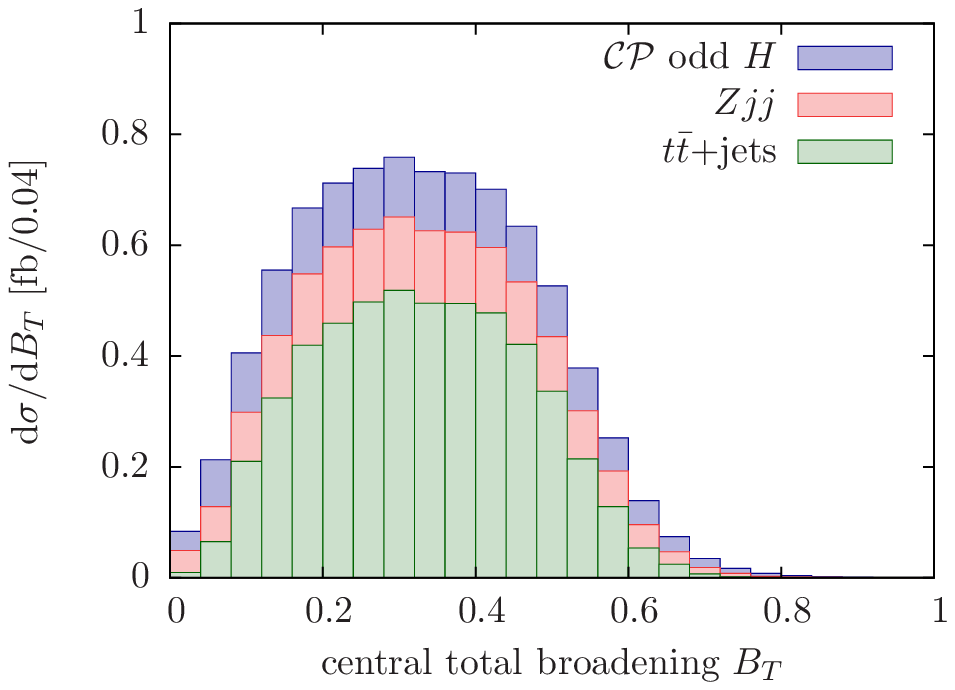}\\[0.5cm]
    \includegraphics[width=0.43\textwidth]{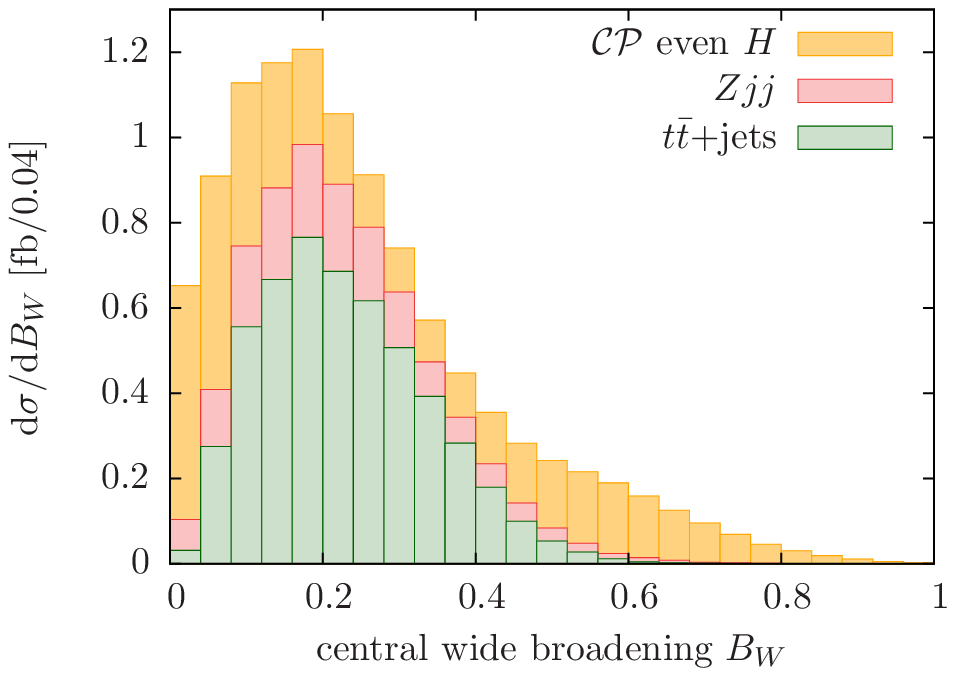}
    \hskip0.6cm
    \includegraphics[width=0.43\textwidth]{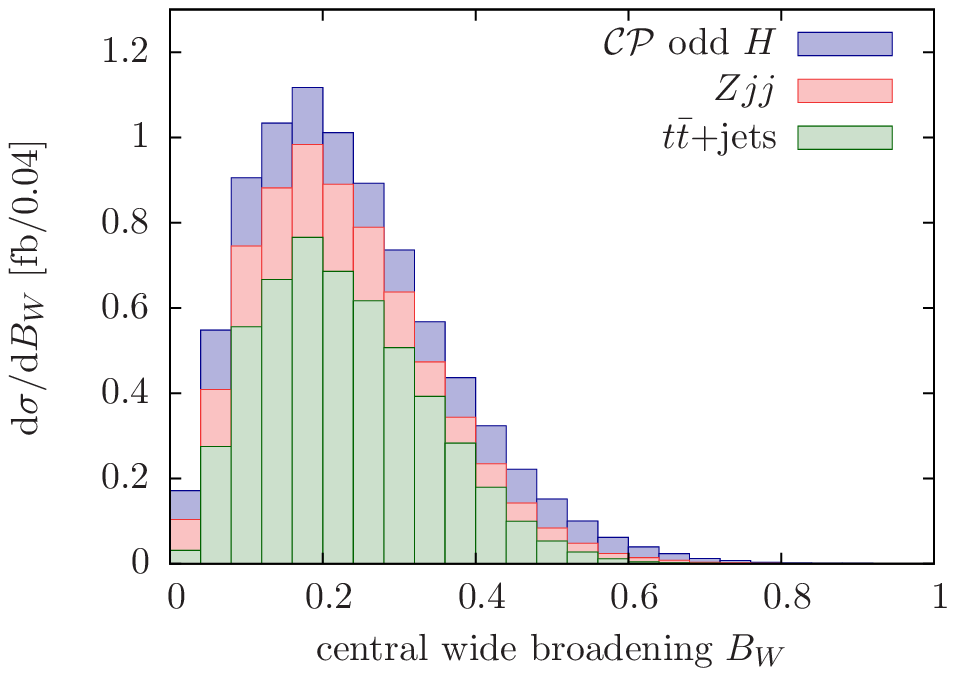}\\[0.5cm]
  \end{center}
  \caption{\label{fig:distribbk} Distributions of the event shape
    observables of Sec.~\ref{sec:shapes} including the background
    after the cuts of Sec.~\ref{sec:selection}.}
\end{figure*}

\begin{subequations}
  \label{eq:selectionb}
  The reconstructed taus need to be hard and central to guarantee a
  good reconstruction efficiency
  \begin{equation}
    p_{T,\tau}\geq 20~\gev,~{\text{and}}~|y_\tau|\leq 2.5\,.
  \end{equation}
  As already mentioned we limit ourselves to the clean purely leptonic
  ditau final state in this paper. It is however worth mentioning,
  that the tau reconstruction algorithms show very good reconstruction
  efficiencies also for (semi)hadronic decays
  \cite{exitau,cmstau,Chatrchyan:2012vp}, so that there is good reason
  to believe that our results can be significantly improved in a more
  realistic analysis.

  The Higgs decay products are required to reconstruct the Higgs mass
  within a $40~\gev$ window,
  \begin{equation}
    \label{eq:massrec}
    |m_{\tau\tau}-m_H|<20~\gev \,,
  \end{equation}
  and the Higgs has to fall between two reconstructed jets,
  \begin{equation}
    \exists\, j_a,j_b: y_{j_a}<y_h<y_{j_b}\,.
  \end{equation}
\end{subequations}
Since we do not consider a full tau reconstruction, the
region defined by Eq.~\gl{eq:massrec} without any further selection
criteria contains all signal events.

If an event passes the above selection criteria, we isolate the Higgs
decay products from the event and feed all remaining final state
particles with $|\eta_i|\leq 4.5$ and $p_{T,i}\geq 1~\gev$ into the
computation of the event shape observables discussed in the previous
section. We therefore implicitly assume that the resonance has already
been established and that the $\tau$ reconstruction is efficient
enough to avoid a large pollution from mistags and/or fakes. A
cut-flow of the analysis steps
Eqs.~\gl{eq:selectiona}-\gl{eq:selectionb} is listed in
Tab.~\ref{tab:cut-flow}. We also include a rough estimate of the
background rejection due to $b$-vetos following by applying a flat
combined $b$ tagging efficiency of $80\%$~\cite{atltag}. We neglect
the signal reduction effect by the mis-tag for illustration purposes
since it is not too large, at most a few percent~\cite{atltag}.
Note that there is good agreement with the results of
Ref.~\cite{Plehn:1999xi}. Note also that the specific selection
criteria that are necessary to reduce the backgrounds can complicate
the resummation of the event shapes. In particular the invariant mass
cuts introduce additional scales to the problem and will have an
impact in the reduction on the theoretical uncertainties.

In order to study the sensitivity of these observables without
introducing a bias, we do not impose a central jet veto
\cite{early,vetos,higgsscale,Rainwater:1998kj,Plehn:1999xi}.  In
Ref.~\cite{Cox:2010ug} it was shown that different cut efficiencies of
jet vetos for WBF and GF contributions can be used to separate WBF
from GF. Therefore, jet vetos in fact provide an ``orthogonal''
strategy to ours. Given that systematic and theoretical uncertainties
of both strategies are different, a comparison or a combination of
both strategies can help to reduce systematics in separating GF from
WBF. This can eventually lead to smaller uncertainties in the
extraction of the Higgs couplings along the lines of Eq.~\gl{eq:1}.

\section{Results}
\label{sec:results}
\subsection{$\cal{CP}$ even vs. $\cal{CP}$ odd}
\label{sec:cpecpo}
We are now ready to study the sensitivity of the shape observables of
Sec.~\ref{sec:shapes} quantitatively. Imposing the selection cuts of
the previous section, we show normalized signal distributions in
Fig.~\ref{fig:distrib} for the ${\cal{CP}}$ even and odd Higgs
cases. As done in Refs.~\cite{vera,Klamke:2007cu,Andersen:2008} we
consider $\Delta\Phi_{jj} \in [-\pi,\pi]$. 

Fig.~\ref{fig:distrib} reveals a substantial dependence on the
${\cal{CP}}$ quantum numbers of the Higgs and the sensitivity in the
azimuthal angle correlation carries over to the event shapes. This is
evident when comparing to, {\emph{e.g.}}, thrust, Eq.~\gl{eq:thrust}:
a \cp~even $Hjj$ event has tagging jets which are preferably
back-to-back. Given that the tagging jets are by construction the
leading jets in the event, we observe a more pencil-like structure for
the thrust observable in $Hjj$ than we see in the ${\cal{CP}}$ odd
$Ajj$ case. In this context, the thrust--$\Delta\Phi_{jj}$ correlation
is particularly interesting,~Fig.~\ref{fig:corr}. Indeed, thrust and
$\Delta\Phi_{jj}$ are fairly correlated as expected after the above
points. This also means that it should be possible to carry over
theoretical and experimental improvements of either observable to the
other one.

\begin{figure}[t]
  \begin{center}
    \includegraphics[width=0.43\textwidth]{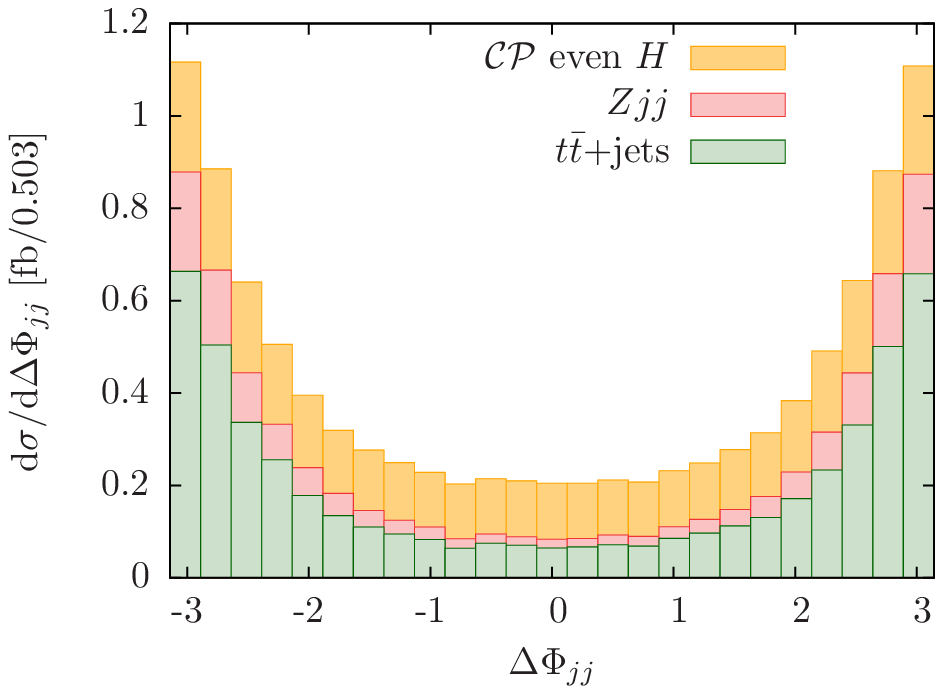}\\[0.5cm]
    \includegraphics[width=0.43\textwidth]{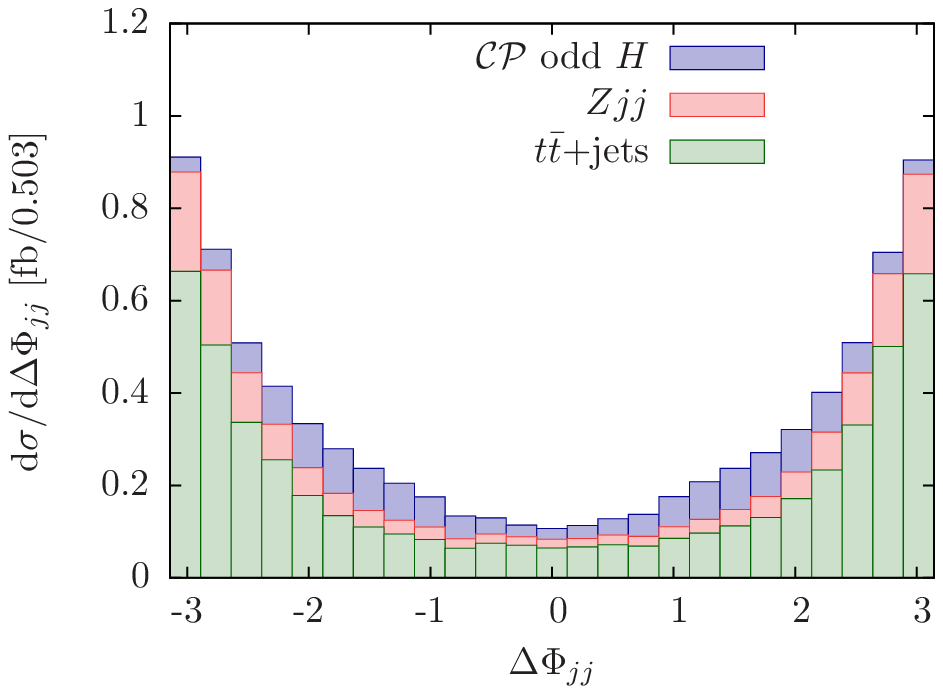}
  \end{center}
  \caption{\label{fig:distribp} $\Delta\Phi_{jj}$ distribution
    including the background after the cuts of
    Sec.~\ref{sec:selection}.}
\end{figure}
\begin{figure}[b]
  \begin{center}
    \includegraphics[width=0.46\textwidth]{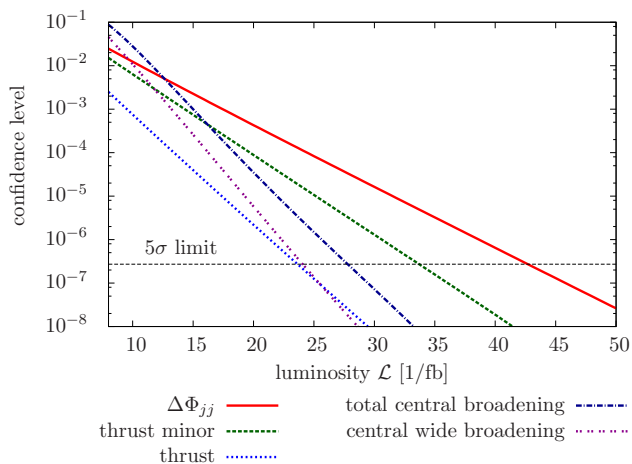}
    \caption{\label{fig:cls} Sensitivity of a binned log-likelihood
      shape comparison of the observables of Figs.~\ref{fig:distribbk}
      and \ref{fig:distribp}. The dotted line corresponds to a
      $5\sigma$ ($2.72\cdot 10^{-7}$ confidence level)
      discrimination.}
  \end{center}
\end{figure}

Another way to understand the special relation of thrust and
$\Delta\Phi_{jj}$ from a different vantage point is by investigating
the jet emission pattern of Higgs+2 jets events.  Due to the observed
Poisson-like scaling pattern in the exclusive number of central
non-tagging jets in Higgs+2 jets events once the cuts of
Sec.~\ref{sec:selection} are applied
\cite{higgsscale,poisson,Rainwater:1998kj,vetos,Plehn:1999xi}, the
two-jet topology plays a special role. The two jets recoiling against
the Higgs therefore largely determine the orientation of the thrust
axis, and, given that both observables are defined in the
beam-transverse plane, we observe a direct connection of thrust with
$\Delta\Phi_{jj}$. This, however, is affected and washed out by soft
radiation (non-resolved jets) included in the first
observable. Suppressing the latter by admitting a more accentuated
role to the two tagging jets, when turning to, {\emph{e.g.}}, cone
thrust minor, we see a more direct correlation with $\Delta\Phi_{jj}$,
Fig.~\ref{fig:corrcone}.

\bigskip

Fig.~\ref{fig:distrib} gives, of course, a wrong impression of the
eventual discriminative power as the normalization relative to the
background and the backgrounds' shape are not
included. Fig.~\ref{fig:distribbk} draws a more realistic picture by
comparing the differential cross sections of the $Ajj$+background and
the $Hjj$+background. In particular, the background mimics the
$\Delta\Phi_{jj}$ distribution of the \cp~even $Hjj$ events and most
of the discriminating power comes from a critical $S/B$. Systematic
uncertainties can easily wash out the small excess around
$|\Delta\Phi_{jj}|\simeq 2$ for $Ajj$ production in comparison to
$Hjj$. A more quantitative statement, however, requires a dedicated
Monte Carlo analysis taking into account experimental systematics and
we cannot explore this direction in our analysis {\emph{in
    extenso}}. The broadening observables, on the other hand, lift the
$\Delta\Phi_{jj}$ signal-background shape-degeneracy especially in the
\cp-even Higgs case.

We perform a binned log-likelihood hypothesis test as considered in
Refs.~\cite{lykken,llhr} to provide a statistically well-defined
estimate of when we will be able to tell apart the \cp~quantum numbers
of a $125~\gev$ SM-like Higgs resonance. At the same time this
provides a statistically well-defined picture of which observable is
particularly suited for this purpose.  Shape differences and different
normalizations ({\emph{i.e}} due to the missing WBF component in $Ajj$
production) are incorporated simultaneously in this approach. We
comment on the discriminative power that solely arises from the
different shapes later in Sec.~\ref{sec:HLL}. In performing the
hypothesis test we treat each individual bin in Fig.~\ref{fig:distrib}
as a counting experiment. Thereby we do not include any shape
uncertainties, which can be different for each of the considered
observables.

Hence, some words of caution are in place. On the one hand,
sensitivity from {\emph{e.g.}} soft radiation pattern that contributes
to the overall sensitivity of the event shape observables can be
weakened by pile-up ({\emph{cf.}}  Sec.~\ref{sec:pile}). On the other
hand, increasing $S/B$ to enhance sensitivity in $\Delta\Phi_{jj}$
heavily relies on jet vetos which can be theoretically challenging.
Also, the experimental resolution (which should be reflected by the
binning in Figs.~\ref{fig:distribbk} and~\ref{fig:distribp}) is
currently not known.

We plot the confidence levels obtained from the hypothesis test in
Fig.~\ref{fig:cls} as a function of the integrated luminosity. When
the confidence level ({\emph{i.e.}} the probability of one hypothesis
to fake the other one) is smaller than $2.72\,\cdot\, 10^{-7}$ one speaks
of a $5\sigma$ discrimination, implicitly assuming Gaussian-like
probability density functions. We see from Fig.~\ref{fig:cls} that
event shapes indeed provide a well-suited class of \cp~discriminating
observables, superseding $\Delta\Phi_{jj}$ within the limitations of
our analysis mentioned above. Fig.~\ref{fig:cls} strongly suggests
that event shape observables should be added to the list of
\cp-sensitive observables which need to be studied at the LHC to
measure the Higgs' \cp.

\subsection{Higgs-lookalike $\cal{CP}$ odd}
\label{sec:HLL}
\begin{figure}[!t]
  \begin{center}
    \includegraphics[width=0.48\textwidth]{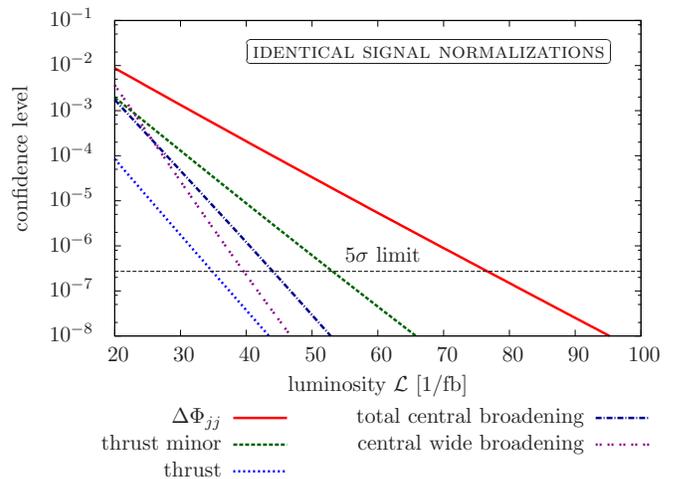}
    \caption{\label{fig:clsll} Sensitivity of a binned log-likelihood
      shape comparison of the observables of Figs.~\ref{fig:distribbk}
      and \ref{fig:distribp} and identically chosen signal
      normalizations according to $Hjj$,~Tab.~\ref{tab:cut-flow}. The
      dotted line corresponds to a $5\sigma$ ($2.72\cdot 10^{-7}$
      confidence level) discrimination.}
    \vspace{-0.4cm}
  \end{center}
\end{figure}

In fact, Fig.~\ref{fig:cls} being the result of a comparison that
reflects both different shape {\emph{and}} normalization of the $Ajj$
and $Hjj$ samples, the sensitivity that arises only due to shape
differences ({\emph{cf.}}  Fig.~\ref{fig:distrib}) is not
obvious. Also, from a phenomenological point of view (and this was one
of our assumptions in Sec.~\ref{sec:selection}), the resonance will
have been discovered before we address its spin and \cp. Therefore the
normalization of the signal will be extracted from data, and only the
subsequent measurement of shapes will be used to extract information
on spin and \cp. Hence, it is reasonable to study the discriminative
power of the event shapes in comparison to $\Delta\Phi_{jj}$ when the
overall normalization after cuts of pseudoscalar and scalar are
identical. This is plotted in Fig.~\ref{fig:clsll}. Again we see that
the event shape observables are good discriminators (the comments of
the previous section are applicable here as well). This also tells us
that a significant share of the discriminative power found in the
previous section stems from the distributions' shape. Especially the
jet broadenings, which exhibit a different background distribution
compared to signal for $Hjj$ as opposed to $\Delta\Phi_{jj}$, should
therefore be stressed as a discriminative observable when considering
systematics.

\begin{figure*}[tb!]
  \begin{center}
    \includegraphics[width=0.41\textwidth]{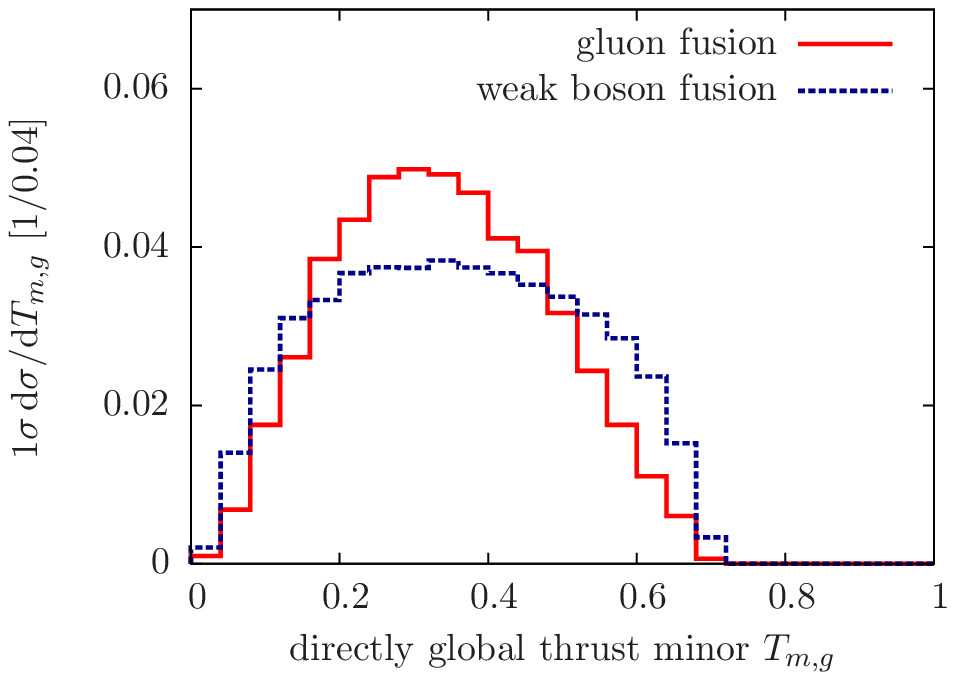}
    \hskip0.6cm
    \includegraphics[width=0.41\textwidth]{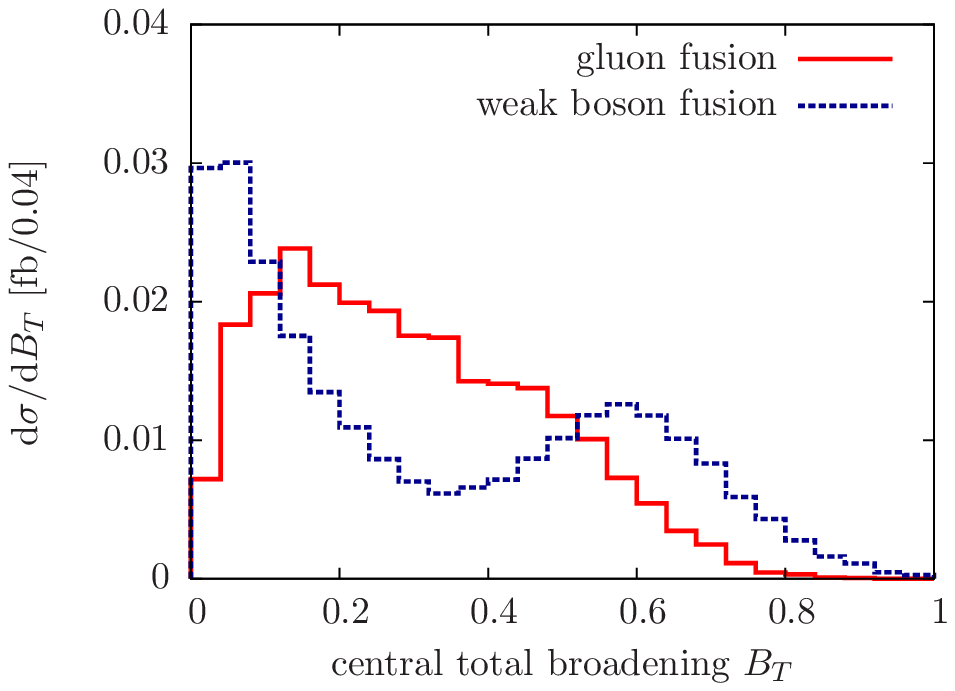}\\[0.5cm]
    \includegraphics[width=0.41\textwidth]{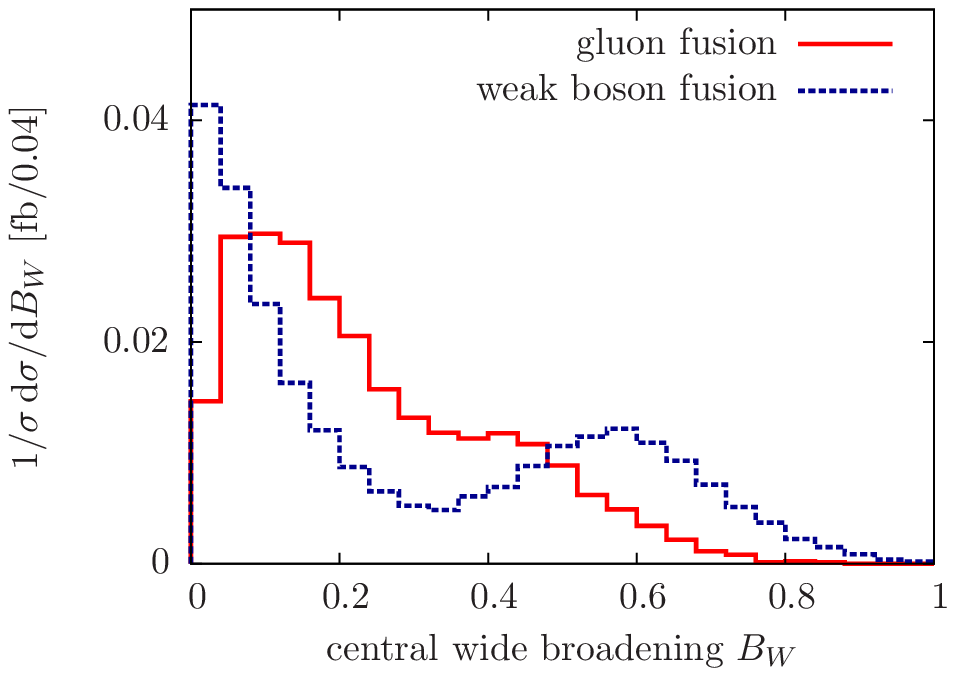}
    \hskip0.6cm
    \includegraphics[width=0.41\textwidth]{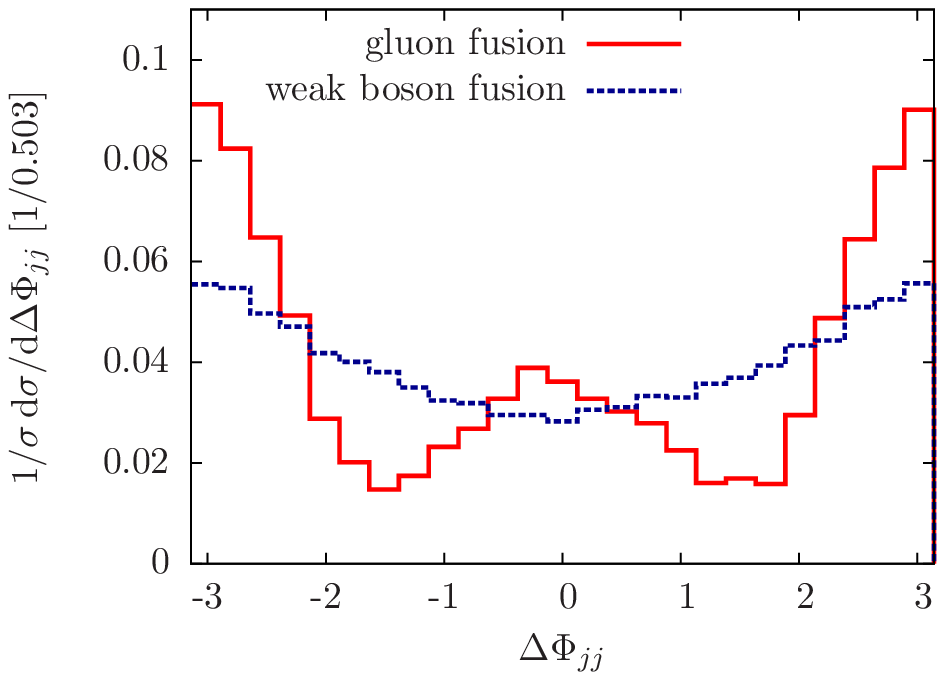}
  \end{center}
  \caption{\label{fig:wbfgf} Normalized distributions of
    $\Delta\Phi_{jj}$ and of the event shape observables of
    Sec.~\ref{sec:shapes} for separate weak boson fusion and gluon
    fusion contributions in case of the \cp~even SM Higgs. The cuts of
    Sec.~\ref{sec:selection} have been applied.}
  \vspace{-0.2cm}
\end{figure*}

\subsection{Toward discriminating gluon fusion and weak boson fusion
  contributions}
Having established the event shape observables as \cp-discriminating
quantities, we move on and discuss the potential of these observables
to help separating WBF from GF, hence contributing to more precise
determination of the Higgs couplings according to Eq.~\gl{eq:1}. We
show normalized signal distributions for the individual WBF and GF
contributions in Fig.~\ref{fig:wbfgf} and we see a similar behavior as
encountered in Fig.~\ref{fig:distrib}.

It is known that unless we include a non-renormalizable $SU(2)_L$
axion-type dimension 5 operator \hbox{$\sim HW\widetilde W$}, where
$\widetilde W$ is the dual $SU(2)_L$ field strength (which
also arises in the SM at the loop level similar to
Eq.~\gl{eq:efflag}), the $\Delta\Phi_{jj}$ distribution is almost
flat in WBF \cite{Plehn:2001nj}. While such an operator should be
constrained experimentally, a sizeable \cp-violating coupling is not
expected from a theoretical perspective. Actually, the strategy
outlined in Secs.~\ref{sec:selection},~\ref{sec:cpecpo} and
\ref{sec:HLL} does not suffer from drawbacks when including explicit
\cp~violation in the gauge sector and remains applicable in a
straightforward way.  In fact, the relative contribution of WBF and GF
to the cross section heavily influences the quantities
Eqs.~\gl{eq:thrust}-\gl{eq:wrapbroad}, and therefore drives the
observed sensitivity in the context of \cp~analyses,
Fig.~\ref{fig:clsll}.

\begin{figure}[t]
  \begin{center}
    \includegraphics[width=0.43\textwidth]{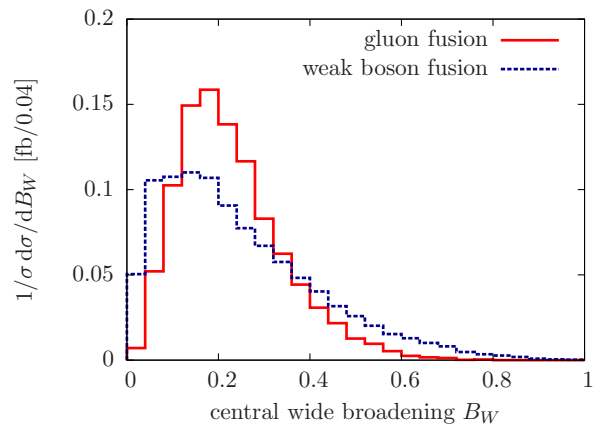}
    \caption{\label{fig:inner} Comparison of the wide broadening for
      the tracks which are not part of the tagging jets for WBF and
      GF.}
    \vspace{-0.4cm}
  \end{center}
\end{figure}

\begin{figure*}[t]
  \begin{center}
    \includegraphics[width=0.43\textwidth]{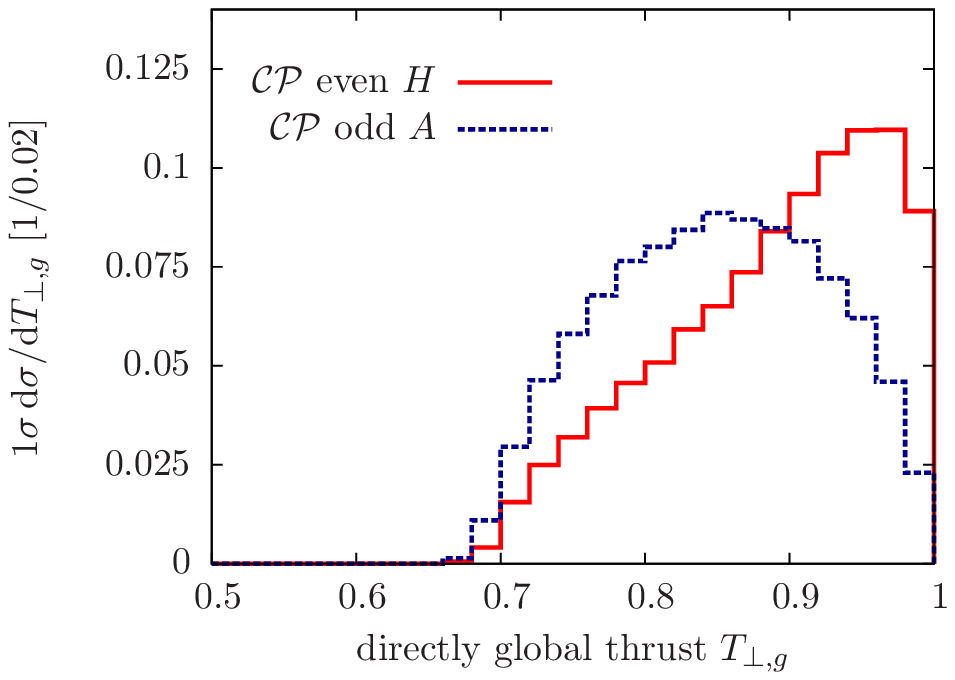}
    \hskip0.6cm
    \includegraphics[width=0.43\textwidth]{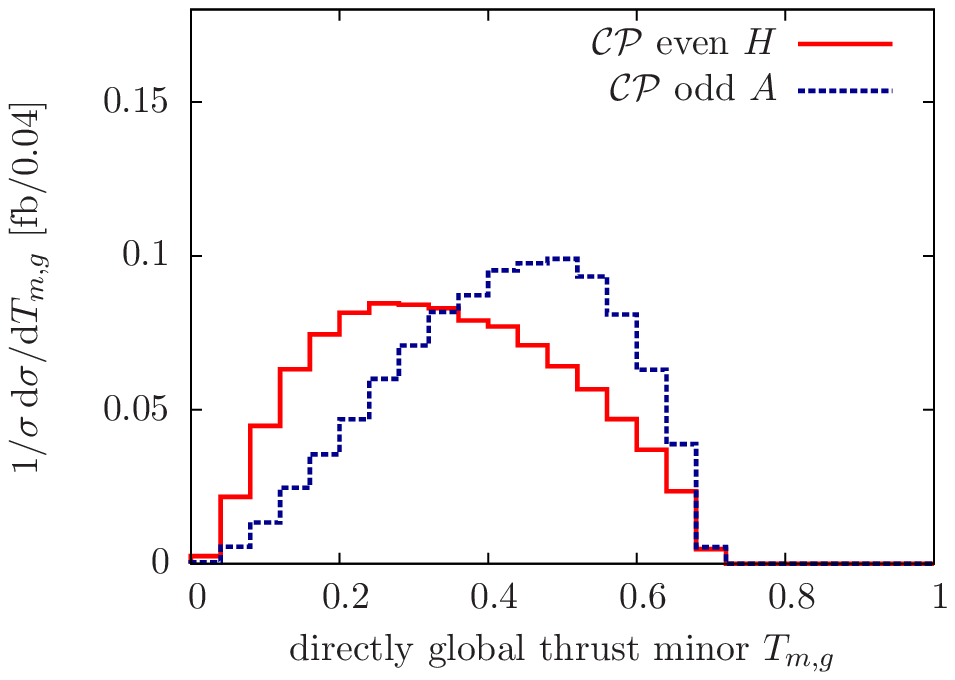}\\[0.5cm]
    \includegraphics[width=0.43\textwidth]{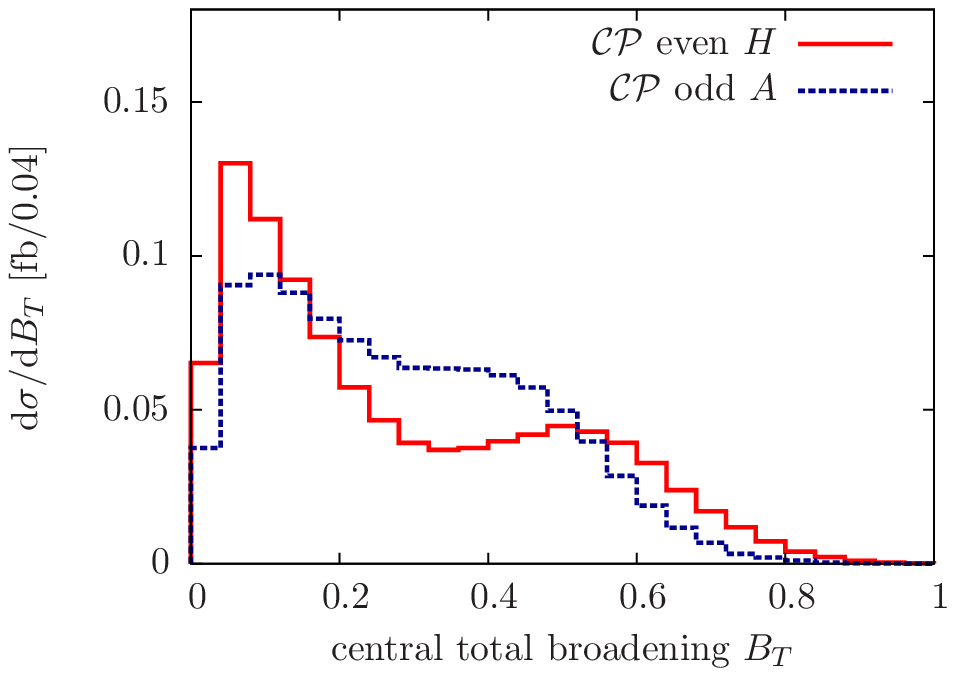}
    \hskip0.6cm
    \includegraphics[width=0.43\textwidth]{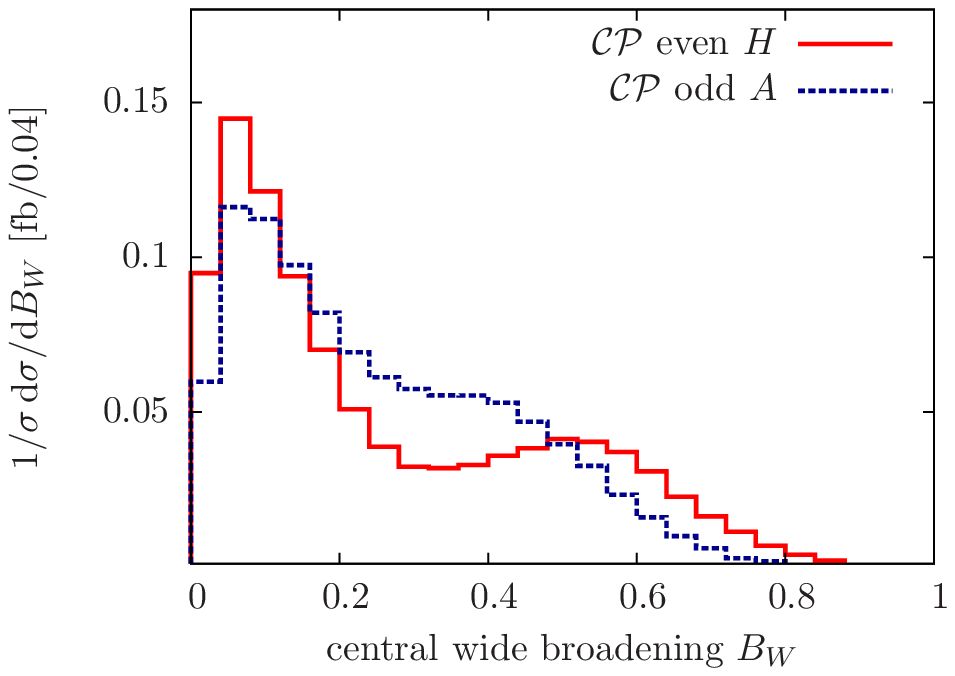}
  \end{center}
  \caption{\label{fig:distribpu} Event shape observables computed from
    the jet constituents as outlined in Sec.~\ref{sec:pile}.}
  \vspace{-0.2cm}
\end{figure*}

Keeping that in mind, we can use the correlations observed in
Fig.~\ref{fig:wbfgf} to separate GF from WBF. There is no meaning in
performing a hypothesis test, so we limit ourselves to a discussion of
the normalized distributions in the following. Forming ratios of
different cut-scenarios in an ABCD-type approach, {\emph{e.g.}}
comparing $0.1\leq B_{W}\leq 0.5$ with the complementary region in a
background-subtracted sample allows to extract the WBF and GF
contributions (we stress again that interference is negligible for the
chosen cuts). An assessment of the uncertainty of such an extraction,
however, requires a realistic simulation, taking into account
experimental systematics, and is beyond the scope of our work.

\subsection{Impact of pile-up}
\label{sec:pile}
A potential drawback, which has not been discussed in depth so far,
arises from the unexpectedly high pile-up activity reported by both
{\sc{Cms}} and {\sc{Atlas}} for the 2011 run. Because soft tracks
enter the evaluation of the event shape observables, which contain
information about \cp~or WBF vs. GF, ({\emph{cf.}}
Fig.~\ref{fig:inner}), we expect pile-up to have an impact on the
event shape phenomenology. Especially in the forward region of the
detector pile-up subtraction is not available. A way to weaken the
phenomenological impact of pile-up is to use jet constituents as input
for the even shape observables. This can distort many of the
theoretical properties of event shapes (in particular resummation
becomes more involved due to introduction of new scales to the
problem). Hence, the potential theoretical improvements are bound to
the experimental capabilities to subtract or reduce pile-up by the
time the resonance is established.

To understanding how much our
sensitivity decreases by using the reconstructed jets' constituents
instead of all particles, we analyze the event shapes again for a
modified cut set up. We stick to the selection criteria
Eqs.~\gl{eq:selectiona1}-\gl{eq:selectionb}, but modify our jet
pre-selection. Again we cluster anti-kT jets with $D=0.4$ but consider
jets
\begin{equation*}
  \label{eq:mod}
  \tag{6a'} 
  \begin{split}
    p_{T,j}\geq 40~\gev\,,\quad&\text{if}~2.5\leq |y_j|\leq 4.5\,,~\text{and} \\
    p_{T,j}\geq 10~\gev\,,\quad&\text{if}~|y_j|\leq 2.5 \,.
  \end{split}
\end{equation*}
In the central region $|y|<2.5$ the tracker can be used to infer the
number of primary vertices of the event and here tracking serves as an
efficient handle to reduce pile-up. In the forward region $|y|>2.5$
pile-up subtraction strategies are scarce and we rely exclusively on
the hardness of the tagging jets to suppress pile-up.

Thus, we require at least three jets in the event, while the hardest
two jets still have to obey $m_{jj}>600~\mathrm{GeV}$, {\emph{i.e.}}
we try to keep as much soft central sensitivity in the first place
({\emph{cf.}}  Fig.~\ref{fig:inner}). Instead of feeding all particles
into the computation of the event shapes, we only take the
constituents of the jets which pass these criteria. Since the
event shapes are weighted in $p_T$ we could also use the recombined
jet four momenta. We found that this will not affect the sensitivity negatively. This
would be the method of choice when facing extreme pile-up
conditions.

The signal cross sections due to the modified selection criteria
decreases to $1.89$ ($1.35$) for $Hjj$ ($Ajj$) production, yielding
$S/B\simeq 0.27~(0.19)$. The result is plotted in
Fig.~\ref{fig:distribpu}. We see that some discriminative power is
lost, but the distributions are still sensitive enough to guarantee
discrimination between \cp~even and odd (and between WBF and GF) at a
however larger integrated luminosity.


\section{Conclusions and Outlook}
\label{sec:conc}
Following the discovery of a new resonance at the LHC, the
determination of its \cp~quantum numbers and its couplings to
SM fermions and gauge bosons will contribute to a more precise
understanding of particle physics at a new energy frontier. Addressing
these questions also poses an important test of the validity of the
Standard Model after the Higgs-like resonance is established.

In this paper we have analyzed the potential of event shape
observables to discriminate between different \cp~hypotheses once a
resonance is established. While more work from both theoretical and
experimental sides is needed, we find excellent discrimination power
for Higgs masses in the vicinity of where {\sc{Atlas}} and {\sc{CMS}}
have reported an excess. Sensitivity in \cp~studies is inherited from
sensitivity in telling apart weak boson fusion and gluon fusion
contributions, making event shape observables natural candidates to
serve this purpose in a realistic experimental analysis. The ability
to separate GF and WBF induced Higgs production will allow for an
improved measurement of the Higgs couplings to fermions and
electroweak gauge bosons.

We find the discriminative power of the event shapes to be fairly
robust when turning to the jet level, and we therefore conclude that
discriminative power should persist even under busy pile-up
conditions. The cuts we choose in Sec.~\ref{sec:selection} to
arrive at these results will be subject to modifications in the
actual experimental analysis and only approximate the realistic
situation, especially when semi-hadronic $H\to \tau\tau$ decays will
be analyzed analogous to Ref.~\cite{Chatrchyan:2012vp}. It is,
however, also clear that realistic selection criteria will affect
the considered observables in a similar fashion. The Higgs decay
products do not explicitly enter our analysis apart from the
reconstruction. Hence modifications of $S/B$ will shift our findings
of Figs.~\ref{fig:cls} and \ref{fig:clsll} to higher absolute
luminosities but will not change the relative improvement of the
observables of Sec.~\ref{sec:shapes} over $\Delta\Phi_{jj}$ (bear in
mind our discussion of pile-up in Sec.~\ref{sec:pile}).

We have limited our analysis to $\sqrt{s}=14$ TeV. Confronting data
with the question for \cp~with a statistically reasonable outcome is
driven by shape analyses which typically involve
${\cal{O}}(50~\ifb)$. It is therefore unlikely that such an analysis
will be performed with the $7~\tev$ or $8~\tev$ run unless the
resonance is significantly overproduced relative to the SM
expectation. Nonetheless, event shape strategies can also be
applied for $7~\tev$ or $8~\tev$ center-of-mass energies.

We have not exhausted the list of potentially sensitive event shape
observables in our analysis.  From a statistical point of view, the
cross sections are large enough to eventually allow a two-dimensional
analysis of observables orthogonal to the event shapes or a
combination with other observables in a neural-net-based
analysis. This should eventually allow to establish the \cp~of a 125
GeV resonance shortly after its discovery.

\bigskip 

{\bf{Acknowledgments}} --- We thank Andrea Banfi, Gavin Salam, and
Giulia Zanderighi for making their {\sc{Caesar}} event shape code
available to us, and we thank especially Gavin Salam for support. We
thank Andy Pilkington for many helpful comments, especially concerning
the separation of gluon and weak boson fusion. We also would like to
thank Tilman Plehn as part of the organizing committee of the
Heidelberg New Physics Forum for creating the environment where the
idea for this work was born.

C.E. acknowledges funding by the Durham International Junior Research
Fellowship scheme. Parts of the simulations underlying this study have
been performed on bwGRiD (\url{http://www.bw-grid.de}), member of the
German D-Grid initiative, funded by the Ministry for Education and
Research (Bundesministerium f\"ur Bildung und Forschung) and the
Ministry for Science, Research and Arts Baden-W\"urttemberg
(Ministerium f\"ur Wissenschaft, Forschung und Kunst
Baden-W\"urttemberg).


\end{document}